\documentclass[%
a4paper,
twocolumn,
aps,
amsmath,
amssymb,
amsfonts,
accepted=2021-09-14]{quantumarticle}
\pdfoutput=1
\usepackage[utf8]{inputenc}
\usepackage[english]{babel}
\usepackage[T1]{fontenc}
\usepackage{
    physics,
    graphicx,
    multirow,
    mathtools, 
    placeins, 
    nicefrac, 
    hyperref 
}
\usepackage{adjustbox}
\usepackage[dvipsnames]{xcolor}
\hypersetup{
    colorlinks,
    linkcolor={blue!50!black},
    citecolor={blue!50!black},
    urlcolor={blue!80!black}
}

\usepackage[justification=justified]{subcaption}
\usepackage[capitalize]{cleveref}

\renewcommand{\var}[1]{{\mathrm{Var}\left[#1\right]}}
\newcommand{\cov}[1]{{\mathrm{Cov}\left[#1\right]}}
\newcommand{\mse}[1]{{\mathrm{MSE}\left[#1\right]}}
\newcommand{\bias}[1]{{\mathrm{Bias}\left[#1\right]}}

\DeclareMathOperator*{\argmin}{arg\,min}

\begin{document}
    
    
    \title{Quantum Error Mitigation using Symmetry Expansion}
    
    
    \author{Zhenyu Cai}
    \email{cai.zhenyu.physics@gmail.com}
    \orcid{0000-0001-5659-4301}
    \affiliation{Department of Materials, University of Oxford, Oxford, OX1 3PH, United Kingdom}
    \affiliation{Quantum Motion Technologies Ltd, Nexus, Discovery Way, Leeds, LS2 3AA, United Kingdom}
    
    
    \begin{abstract}
         Even with the recent rapid developments in quantum hardware, noise remains the biggest challenge for the practical applications of any near-term quantum devices. Full quantum error correction cannot be implemented in these devices due to their limited scale. Therefore instead of relying on engineered code symmetry, symmetry verification was developed which uses the inherent symmetry within the physical problem we try to solve. In this article, we develop a general framework named symmetry expansion which provides a wide spectrum of symmetry-based error mitigation schemes beyond symmetry verification, enabling us to achieve different balances between the estimation bias and the sampling cost of the scheme. We show that certain symmetry expansion schemes can achieve a smaller estimation bias than symmetry verification through cancellation between the biases due to the detectable and undetectable noise components. A practical way to search for such a small-bias scheme is introduced. By numerically simulating the Fermi-Hubbard model for energy estimation, the small-bias symmetry expansion we found can achieve an estimation bias $6$ to $9$ times below what is achievable by symmetry verification when the average number of circuit errors is between $1$ to $2$. The corresponding sampling cost for random shot noise reduction is just $2$ to $6$ times higher than symmetry verification. Beyond symmetries inherent to the physical problem, our formalism is also applicable to engineered symmetries. For example, the recent scheme for exponential error suppression using multiple noisy copies of the quantum device is just a special case of symmetry expansion using the permutation symmetry among the copies.
    \end{abstract}
    
    \maketitle

    \section{Introduction}\label{sec:intro}
    The ultimate goal of implementing a fully fault-tolerant quantum error correction scheme for quantum algorithms with provable speed-up may still be years away. However, the recent rapid advance in quantum computing hardware has enabled certain computational tasks to be performed that are well beyond classical capabilities~\cite{aruteQuantumSupremacyUsing2019,zhongQuantumComputationalAdvantage2020}, prompting the question of whether it is possible to perform any practically useful tasks with the noisy intermediate-scale quantum (NISQ) devices that we will soon have. 
    
    One of the biggest roadblocks for such a goal is how to alleviate the damage done by the noise in these NISQ devices with the limited quantum resource we have. This brings us to \emph{quantum error mitigation}, which unlike quantum error correction, mostly relies on performing additional measurements instead of employing additional qubits to mitigate the damages caused by errors~\cite{temmeErrorMitigationShortDepth2017,endoPracticalQuantumError2018,caiMultiexponentialErrorExtrapolation2021, endoHybridQuantumClassicalAlgorithms2021}. Quantum error mitigation has been successfully implemented in various experimental settings~\cite{kandalaErrorMitigationExtends2019, giurgica-tironDigitalZeroNoise2020, laroseMitiqSoftwarePackage2020, googleaiquantumandcollaboratorsHartreeFockSuperconductingQubit2020, zhangErrormitigatedQuantumGates2020}. \emph{Symmetry verification} is one such error-mitigation technique that projects the noisy output quantum state back into the symmetry subspace defined by the physical problem we try to solve~\cite{bonet-monroigLowcostErrorMitigation2018, mcardleErrorMitigatedDigitalQuantum2019}. Besides measuring all symmetries in every circuit run and discarding the runs that fail any symmetry test, symmetry verification can also be carried out in a post-processing way with a much simpler measurement scheme inspired by quantum subspace expansion~\cite{mccleanHybridQuantumclassicalHierarchy2017, bonet-monroigLowcostErrorMitigation2018}. Since stabiliser code decoding can be viewed as symmetry verification with engineered symmetry, we can also apply similar post-processing techniques to reduce the practical challenges in the stabiliser code implementation~\cite{mccleanDecodingQuantumErrors2020}. 
    
    More recently, two teams have independently developed a quantum error mitigation scheme that uses multiple copies of the noisy quantum device to suppress errors exponentially as the number of copies increases~\cite{koczorExponentialErrorSuppression2020,hugginsVirtualDistillationQuantum2021}. The method was named \emph{virtual distillation} by one of the teams. Its core idea relies on the permutation symmetry among the copies, but the way it is implemented does not involve projecting the noisy state into the correct symmetry subspace, and thus does not fall within the symmetry verification framework. In fact, the corresponding permutation symmetry verification~\cite{berthiaumeStabilisationQuantumComputations1994,barencoStabilizationQuantumComputations1997,peresErrorSymmetrizationQuantum1999} can only suppress errors linearly with the increase of the number of copies instead of exponentially. 
    
    In this article, we will provide a general framework named \emph{symmetry expansion} that encompasses a much wider range of symmetry-based error mitigation schemes beyond symmetry verification. We will discuss ways to search within this wide range of symmetry expansion schemes and identify one that can outperform symmetry verification, just as virtual distillation arises from the permutation symmetry group and vastly outperform the permutation symmetry verification.
    
    We will start by introducing the general concepts of quantum error mitigation in \cref{sec:qem} and the framework of symmetry expansion in \cref{sec:sym_expand}. Then in \cref{sec:sym_exp_bias} and \cref{sec:implementation_cost}, we will discuss the estimation bias and the implementation costs for symmetry expansion. Moving onto \cref{sec:find_opt_sym}, we will outline a practical way to identify a good symmetry expansion scheme. In the end, we will finish with an explicit numerical example for the application of symmetry expansion in \cref{sec:simulation}, along with the conclusion and outlook in \cref{sec:concl}.
    
    \section{Quantum Error Mitigation}\label{sec:qem}
    In quantum error correction, we encodes the quantum information into a code space by employing additional qubits, thus we can recover the noiseless state $\rho_0$ by project our noisy state $\rho$ back into the code space. On the other hand, in quantum error mitigation, instead of trying to recover the ideal \emph{state}, we are trying to recover the ideal \emph{measurement statistics} for some observable. For a given observable $O$, we will slightly abuse the notation and denote the random variable corresponding to its measurement results of the \emph{unmitigated noisy state} $\rho$ as simply $O$, i.e. $\expval{O} = \Tr(O\rho)$, while the random variable corresponding to the measurement results of the \emph{noiseless state} $\rho_0$ is denoted as $O_0$, i.e. $\expval{O_0} = \Tr(O\rho_0)$. Using the noisy circuit to estimate the noiseless expectation value can lead to a large \emph{estimation bias} $\abs{\bias{O}} = \abs{\expval{O} - \expval{O_0}}$ when there is a large amount of noise present. Hence, we can instead try to construct an error-mitigated estimator $O_{\mathrm{em}}$ using the noisy data obtained from different noisy circuit configurations, which would hopefully lead to a smaller bias in the estimates:
    \begin{align*}
        \abs{\bias{O_{\mathrm{em}}}} \leq \abs{\bias{O}}.
    \end{align*}
    Note that throughout this article, we often refer to the magnitude of bias simply as ``bias'' just like above, the exact meaning should be clear from the context. 

    An error-mitigated estimator that reduces the bias will usually come at a cost of a larger \emph{variance}
    \begin{align*}
        \var{O_{\mathrm{em}}} \geq \var{O} \sim \var{O_0}
    \end{align*}
    which means that a larger number of samples are needed to reduce the shot noise in the sample estimate. Very often we care more about the factor of increase of variance for a given error mitigation scheme compared to the unmitigated scheme, which will be called its \emph{sampling cost}:
    \begin{align}\label{eqn:sampling_cost}
     C_{\mathrm{em}} = \frac{\var{O_{\mathrm{em}}}}{\var{O}}.
    \end{align}
    Similarly for biases, we often care more about the \emph{fractional} deviation from the ideal value, which we will call the \emph{relative bias} and denoted as:
    \begin{align}\label{eqn:relative_bias}
     \epsilon_{\mathrm{em}}(O) = \frac{\expval{O_0} - \expval{O_{\mathrm{em}}}}{\expval{O_0}} = - \frac{\bias{O_{\mathrm{em}}}}{\expval{O_0}}.
    \end{align}
    The additional minus sign relative to the bias is for a simpler connection to quantum state infidelity later, but the essential idea does not change. 
    
    When trying to compare two different error mitigation schemes, if one of the schemes has both a smaller bias and a smaller variance (sampling cost), then that would simply be the better scheme. However, the more common and more interesting case is when one scheme has a smaller bias while the other has a smaller variance, which means we need to compare the bias-variance trade-off between the two schemes more carefully. As the number of samples increase, the shot noise contribution from the variance will decrease and the error due to the bias will become more and more dominant, and as a result the small-bias scheme would become more and more favourable. In \cref{sec:err_mit_comp}, we have shown that in practice the minimum number of samples required for a small-bias estimator $A$ to outperform a small-variance estimator $B$ for a Pauli observable $O$ is:
    \begin{align}\label{eqn:Nc_relative}
        N^* \lesssim  \frac{C_A -C_B}{\epsilon_{B}(O)^2 - \epsilon_{A}(O)^2}.
    \end{align} 
    
    As we will see later, the general framework of symmetry expansion will enable us to predict the bias and variance of different symmetry-based error-mitigation schemes, and thus enable us to choose a suitable scheme for our given scenario (or at least rule out some impractical schemes). 
     
    \section{Symmetry Expansion}\label{sec:sym_expand}
    We are given a circuit that would ideally generate the noiseless state $\ket{\psi_0}$ which is stabilised by a group of symmetry operations $\mathbb{G}$:
    \begin{align*}
        G\ket{\psi_0} = \ket{\psi_0} \quad \forall G \in \mathbb{G}. 
    \end{align*}
    The group of symmetry is usually identified through our knowledge of the structure of the circuit and/or the physical meaning of the state (see \cref{sec:output_sym}).
    
    The symmetry subspace that $\ket{\psi_0}$ lives in, within which all $G \in \mathbb{G}$ will act trivially, is defined by the projector~\cite{tungGroupTheoryPhysics1985}:
    \begin{align}\label{eqn:proj_op}
        \Pi_{\mathbb{G}} = \frac{1}{\abs{\mathbb{G}}}\sum_{G \in \mathbb{G}} G.
    \end{align}
    By definition we have
    \begin{equation}\label{eqn:proj_property}
        \begin{aligned}
            \Pi_{\mathbb{G}}^2 &= \Pi_{\mathbb{G}},\quad \Pi_{\mathbb{G}} \ket{\psi_0} =  \ket{\psi_0}.
        \end{aligned}
    \end{equation}
    Note that since $\mathbb{G}$ is a unitary group, we know that for any $G \in \mathbb{G}$ we also have $G^\dagger \in \mathbb{G}$. Therefore $\Pi_{\mathbb{G}}$ being an uniform average of elements in $\mathbb{G}$ will be Hermitian: $\Pi_{\mathbb{G}}^\dagger= \Pi_{\mathbb{G}}$.
    
    In reality, the output of the circuit is a noisy state $\rho$ that might have components outside the symmetry subspace. Hence, we will project $\rho$ back into the symmetry subspace to mitigate the damage of the noise and we will call this process \emph{symmetry verification}~\cite{bonet-monroigLowcostErrorMitigation2018, mcardleErrorMitigatedDigitalQuantum2019}. The effective density matrix we obtain after symmetry verification is simply:
    \begin{align}\label{eqn:rho_sym}
        \rho_{\mathrm{sym}} = \frac{ \Pi_{\mathbb{G}} \rho \Pi_{\mathbb{G}}}{\Tr(\Pi_{\mathbb{G}}\rho)} = \frac{ \Pi_{\mathbb{G}} \rho \Pi_{\mathbb{G}}}{\expval{\Pi_{\mathbb{G}}}}.
    \end{align}
    Throughout this article, we will use $\expval{O} = \Tr(O\rho)$ to denote the expectation value of some observable $O$ from the \emph{noisy} circuit. Using \cref{eqn:rho_sym}, the expectation value of some observable $O$ after symmetry verification will be:
    \begin{align}\label{eqn:ver_obs_gen}
        \Tr(O\rho_{\mathrm{sym}}) = \frac{ \Tr(O\Pi_{\mathbb{G}} \rho \Pi_{\mathbb{G}})}{\expval{\Pi_{\mathbb{G}}}} = \frac{\expval{\Pi_{\mathbb{G}}O\Pi_{\mathbb{G}}}}{\expval{\Pi_{\mathbb{G}}}}.
    \end{align}
    
    In \emph{direct symmetry verification}, we will measure the observable $O$ and the symmetry verification result $\Pi_{\mathbb{G}}$ in the same circuit run, and discard the circuit runs that has $\Pi_{\mathbb{G}}$ measured to be $0$. This would require
    \begin{align}\label{eqn:o_pi_commute}
        \left[O, \Pi_{\mathbb{G}}\right] = 0
    \end{align} 
    and thus we have:
    \begin{align}\label{eqn:ver_obs}
        \Tr(O\rho_{\mathrm{sym}}) = \frac{ \Tr(O\Pi_{\mathbb{G}} \rho )}{\expval{\Pi_{\mathbb{G}}}} = \frac{\expval{O\Pi_{\mathbb{G}}}}{\expval{\Pi_{\mathbb{G}}}}.
    \end{align}
    Alternatively, \cref{eqn:ver_obs} will also hold if we have
    \begin{align}\label{eqn:rho_pi_commute}
        \left[\rho, \Pi_{\mathbb{G}}\right] = 0.
    \end{align}
    How this may arise will be further discussed in \cref{sec:reduce_search_space}. Note that the normalisation factor $\expval{\Pi_{\mathbb{G}}}$ is essentially the fraction of circuit runs that pass the symmetry verification (i.e. those runs with $\Pi_{\mathbb{G}}$ measured to be $1$). 
    
    We can also apply symmetry verification in a \emph{post-processing} way. Using \cref{eqn:proj_op}, we see that we can actually obtain $\expval{\Pi_{\mathbb{G}}}$ by uniformly sampling random $G \in \mathbb{G}$. Similarly, we can also obtain $\expval{O\Pi_{\mathbb{G}}}$ by uniformly sampling random $OG$ for $G \in \mathbb{G}$. In each circuit run, if we measure $O$ along with a random $G \in \mathbb{G}$, we can then compose them to obtain both $G$ and $OG$ and thus effectively obtaining samples for $\expval{\Pi_{\mathbb{G}}}$ and $\expval{O\Pi_{\mathbb{G}}}$, which in turn can give us the estimate of $\Tr(O\rho_{\mathrm{sym}})$ via \cref{eqn:ver_obs}.
    
    Instead of sampling uniformly from $\mathbb{G}$, we can sample each symmetry $G$ with a \emph{positive} weight $w_G$, and we will call this more general sampling scheme \emph{symmetry expansion}. By denoting the sampling weight distribution as $\vec{w}$, the resultant observable following our new sampling scheme will be:
    \begin{align}\label{eqn:exp_obs}
        \Tr(O\rho_{\vec{w}}) = \frac{ \Tr(O\Gamma_{\vec{w}} \rho )}{\Tr(\Gamma_{\vec{w}}\rho )} = \frac{\expval{O\Gamma_{\vec{w}}}}{\expval{\Gamma_{\vec{w}}}}
    \end{align}
    where $\Gamma_{\vec{w}}$ is the \emph{symmetry expansion operator}:
    \begin{align}\label{eqn:exp_op}
        \Gamma_{\vec{w}} = \frac{\sum_{G \in \mathbb{G}} w_G G}{\sum_{G \in \mathbb{G}} w_G} 
    \end{align}
    and $\rho_{\vec{w}}$ is the symmetry-expanded `density operator':
    \begin{align}\label{eqn:rho_exp}
        \rho_{\vec{w}} = \frac{\Gamma_{\vec{w}}\rho}{\Tr(\Gamma_{\vec{w}} \rho )} = \frac{\Gamma_{\vec{w}}\rho}{\expval{\Gamma_{\vec{w}}} }.
    \end{align}
    The symmetry expansion operation acting on $\rho$ here is not necessarily a positive map. Hence, though $\rho_{\vec{w}}$ has unit trace as expected, it is not necessarily positive semi-definite and thus is not necessarily a valid density operator in the usual sense. 
    Note that $\Gamma_{\vec{w}}$ is also a symmetry operator:
    \begin{align}\label{eqn:exp_op_sym}
        \Gamma_{\vec{w}}  \ket{\psi_0} = \ket{\psi_0},\quad \Gamma_{\vec{w}}\Pi_{\mathbb{G}} = \Pi_{\mathbb{G}}.
    \end{align} 
    
    An important class of symmetry expansion would be a \emph{uniform} expansion using a subset of symmetry operators $\mathbb{F} \subseteq \mathbb{G}$ (note that $\mathbb{F}$ is not necessarily a group), which would be our main focus later on. We will denote the corresponding weight vector as $\vec{w} = \vec{\mathbb{F}}$ and the symmetry expansion operator is simply
    \begin{align*}
        \Gamma_{\vec{\mathbb{F}}} = \frac{1}{\abs{\mathbb{F}}}\sum_{F \in \mathbb{F}} F.
    \end{align*}
    When $\mathbb{F}$ contains only one symmetry operator $F$, we will slightly abuse the notation and write   $\mathbb{F} = F$ and thus $\Gamma_{\vec{F}} \equiv F$.
    
    The unmitigated noisy state is simply a special case of symmetry expansion using just the identity operator:
    \begin{align*}
        \Gamma_{\vec{I}}&=I, \quad \rho_{\vec{I}} \equiv \rho.
    \end{align*} 
    Symmetry verification, when used for the purpose of expectation value estimation in \cref{eqn:ver_obs}, is equivalent to a uniform symmetry expansion over the full set of symmetry operators $\mathbb{G}$:
    \begin{align}\label{eqn:sym_new_note}
        \Gamma_{\vec{\mathbb{G}}} & = \Pi_{\mathbb{G}}, \quad \rho_{\vec{\mathbb{G}}} \equiv \rho_{\mathrm{sym}}.
    \end{align}
    Hence, from here on, our discussion of general symmetry expansion schemes will also be applicable to
    \begin{itemize}
        \item the unmitigated case:  $\vec{w} = \vec{I}$,
        \item the symmetry-verified case: $\vec{w} = \vec{\mathbb{G}}$.
    \end{itemize}
    
    As discussed in Refs.~\cite{bonet-monroigLowcostErrorMitigation2018}, symmetry verification is just a special case of \emph{quantum subspace expansion}~\cite{mccleanHybridQuantumclassicalHierarchy2017} and the idea of changing the weight of the symmetry in the projection operator has been discussed before under the quantum subspace expansion framework~\cite{mccleanDecodingQuantumErrors2020}. Assuming all symmetry operators $G \in \mathbb{G}$ commute with the observable $O$ or the noisy state $\rho$ (a special case of \cref{eqn:o_pi_commute} and \cref{eqn:rho_pi_commute}), the expectation value after performing quantum subspace expansion using the expansion operator $\Gamma_{\vec{v}}$ is given by 
    \begin{align}\label{eqn:sub_exp_obs}
        \frac{ \Tr(O\Gamma_{\vec{v}} \rho \Gamma_{\vec{v}}^\dagger)}{\Tr(\Gamma_{\vec{v}}\rho \Gamma_{\vec{v}}^\dagger)} = \frac{\expval{ O\Gamma_{\vec{v}}^\dagger\Gamma_{\vec{v}}}}{\expval{\Gamma_{\vec{v}}^\dagger\Gamma_{\vec{v}}}}.
    \end{align}
    Such quantum subspace expansion using symmetry operators will simply be called symmetry \emph{subspace} expansion (not to be confused with symmetry expansion). Compared to \cref{eqn:exp_obs}, we see that symmetry \emph{subspace} expansion is simply the special case of symmetry expansion in which the expansion operator $\Gamma_{\vec{w}}$ is the square of some other expansion operator: $\Gamma_{\vec{w}} = \Gamma_{\vec{v}}^\dagger \Gamma_{\vec{v}}$, i.e. $\Gamma_{\vec{w}}$ is positive semi-definite. In \cref{eqn:opt_sub_exp}, we have shown that the symmetry \emph{subspace} expansion scheme that would maximise the fidelity against the ideal state is simply the symmetry verification scheme. Non-uniform weight distribution for symmetry \emph{subspace} expansion will only bring advantages over symmetry verification when we are restricted to a truncated set of symmetry operators that do not form a group~\cite{mccleanDecodingQuantumErrors2020}. Hence, in this article we will only focus on the comparison between symmetry verification and symmetry expansion since we are interested in the case in which we have access to the full set of symmetries.
    
    To perform direct symmetry verification for a general symmetry group, we can have an ancilla register with its different basis states corresponding to different elements in the group, then perform control-symmetry operations from the ancilla to the data register and measure out the ancilla as outlined in Ref.~\cite{barencoStabilizationQuantumComputations1997}. For Pauli symmetry groups (stabiliser groups), their direct symmetry verification can be carried out by simply measuring the group \emph{generators} since they are abelian. Similarly, for post-processing symmetry verification, Pauli symmetries are much easier to measure compared to general symmetries since they are Hermitian. Hence, Pauli symmetries are most relevant to the practical implementations of symmetry-based error mitigation techniques and will be our main focus in this article. As a result, the symmetry expansion operator $\Gamma_{\vec{w}}$ we look at will be Hermitian, and since all the symmetries are unitary, we have
    \begin{align}\label{eqn:sym_expand_bound}
        -1 \leq \expval{\Gamma_{\vec{w}}}  \leq 1.
    \end{align}
    
    Though our discussion mainly focuses on Pauli symmetries, many of the arguments hereafter are still applicable to general symmetries. An example for the application of symmetry expansion beyond Pauli symmetry is the virtual distillation scheme mentioned in \cref{sec:intro}, which makes use of the permutation symmetries among identical copies of the quantum system. The details of how it relates to symmetry expansion are outlined in \cref{sec:virt_distill}. Being a symmetry expansion scheme, it was shown to vastly outperform the corresponding symmetry verification scheme via both theoretical analysis and numerical simulations~\cite{koczorExponentialErrorSuppression2020, hugginsVirtualDistillationQuantum2021}.

    \section{Estimation Bias of Symmetry Expansion}\label{sec:sym_exp_bias}
    \subsection{Absolute Infidelity}\label{sec:good_sym_cond}
    Using \cref{eqn:relative_bias}, the \emph{relative bias} for a given observable $O$ when symmetry expanded with weight $\vec{w}$ is simply:
    \begin{equation}
        \begin{aligned}\label{eqn:frac_err_def}
            \epsilon_{\vec{w}}(O) &= 1 - \frac{\Tr(O\rho_{\vec{w}})}{\Tr(O\rho_0)}.
        \end{aligned}
    \end{equation}
    Here $\rho_0 = \ket{\psi_{0}}\bra{\psi_{0}}$ and $\rho_{\vec{w}}$ are the ideal state and symmetry-expanded state, respectively. Recall that the unmitigated case is simply $\vec{w} = \vec{I}$ while the symmetry-verified case is $\vec{w} = \vec{\mathbb{G}}$. 
    
    To show that symmetry verification can produce a smaller estimation bias than the unmitigated case, one possible way is compare their output state fidelity against the ideal state:
    \begin{align}
        \text{Unmitigated: }&F_{\vec{I}}= \Tr(\rho \rho_0) = \expval{\rho_0}\label{eqn:unmit_fid}\\
        \text{Symmetry-verified: }& F_{\vec{\mathbb{G}}} = \Tr(\rho_{\vec{\mathbb{G}}} \rho_0) = \frac{\expval{\rho_0}}{\expval{\Gamma_{\vec{\mathbb{G}}}}} \label{eqn:sym_fid}.
    \end{align}
    Here we have used \cref{eqn:rho_exp} and \cref{eqn:exp_op_sym}. We see that the symmetry-verified fidelity is just the unmitigated noisy fidelity $\expval{\rho_0}$ boosted by a factor of $\expval{\Gamma_{\vec{\mathbb{G}}}}^{-1}$. 
    
    We might be tempted to generalise the above arguments and try to predict the performance of symmetry expansion using the symmetry-expanded fidelity :
    \begin{align}\label{eqn:exp_fid}
       F_{\vec{w}}  = \Tr(\rho_0\rho_{\vec{w}}) & =  \frac{\expval{\rho_0}}{\expval{\Gamma_{\vec{w}}}},
    \end{align}
    which is just the unmitigated noisy fidelity $\expval{\rho_0}$ boosted by a factor of $\expval{\Gamma_{\vec{w}}}^{-1}$. However, unlike the unmitigated and symmetry-verified cases, $\Tr(\rho_0\rho_{\vec{w}})$ is not upper-bounded by $1$ anymore since $\rho_{\vec{w}}$ might not be positive semi-definite as mentioned in \cref{sec:sym_expand}. Hence, instead we construct a metric that measure how far $F_{\vec{w}} = \Tr(\rho_0\rho_{\vec{w}})$ deviate from the ideal value of $1$, which is equivalent to the magnitude of the bias in \cref{eqn:frac_err_def} with the observable being the ideal state $\rho_0$:
    \begin{align}\label{eqn:abs_infid}
        \abs{\epsilon_{\vec{w}}(\rho_0)} = \abs{1 - F_{\vec{w}}} = \abs{1 - \frac{\expval{\rho_0}}{\expval{\Gamma_{\vec{w}}}}}.
    \end{align}
    For the unmitigated and symmetry-verified cases we have:
    \begin{align}
        \abs{\epsilon_{\vec{I}}(\rho_0)} &= \epsilon_{\vec{I}}(\rho_0)  = 1 - F_{\vec{I}} =  1 - \expval{\rho_0}\label{eqn:noisy_infid}\\
        \abs{\epsilon_{\vec{\mathbb{G}}}(\rho_0)} &= \epsilon_{\vec{\mathbb{G}}}(\rho_0) = 1 - F_{\vec{\mathbb{G}}}= 1 - \frac{\expval{\rho_0}}{\expval{\Gamma_{\vec{\mathbb{G}}}}}.\label{eqn:sym_infid}
    \end{align}
    which are simply the output state infidelities in the corresponding cases. Hence, we will call this metric the \emph{absolute infidelity}.
    
    In this article, we are going to use the absolute infidelity, which is the magnitude of the bias with observable $\rho_0$, as our metric for the magnitude of the bias for a given error mitigation scheme.  
    We have further discussed why this is a reasonable metric in \cref{sec:abs_inf_detail}, and the validity of this metric will also be further verified in our numerical simulation later in \cref{sec:simulation}. 
    
    \subsection{Mechanism of Symmetry Expansion}\label{sec:mech_sym_exp}
    The erroneous component of the symmetry-expanded state $\rho_{\vec{w}}$ is defined to be the components that are orthogonal to the ideal state $\rho_0$, which can be obtained by applying the projector $1 - \rho_0$.  Within these erroneous components, there are components living within the symmetry subspace defined by the projector $\Gamma_{\vec{\mathbb{G}}}$, which are undetectable via symmetry verification. The coefficient of these undetectable erroneous components in $\rho_{\vec{w}}$  is simply:
    \begin{align}\label{eqn:undet_prob}
        B_{u, \vec{w}} = \Tr(\Gamma_{\vec{\mathbb{G}}}(1 - \rho_{0}) \rho_{\vec{w}}) = \frac{\expval{\Gamma_{\vec{\mathbb{G}}}} - \expval{\rho_0}}{\expval{\Gamma_{\vec{w}}}}.
    \end{align}
    On the other hand, the erroneous components living outside the symmetry subspace would be detectable, whose coefficient in $\rho_{\vec{w}}$ is:
    \begin{align}\label{eqn:det_prob}
        B_{d, \vec{w}} = \Tr(\left(1 - \Gamma_{\vec{\mathbb{G}}}\right)(1 - \rho_{0}) \rho_{\vec{w}}) = \frac{\expval{\Gamma_{\vec{w}}} - \expval{\Gamma_{\vec{\mathbb{G}}}}}{\expval{\Gamma_{\vec{w}}}}.
    \end{align}
    From \cref{eqn:abs_infid}, we see that the absolute infidelity is simply the magnitude of the sum of the detectable and undetectable error coefficients:
    \begin{align}\label{eqn:sum_of_bias}
        \abs{\epsilon_{\vec{w}}(\rho_0)} = \abs{B_{u, \vec{w}} + B_{d, \vec{w}}}.
    \end{align}
    Hence, $B_{d, \vec{w}}$ and $B_{u, \vec{w}}$ are simply the bias of the symmetry-expanded estimation of the observable $\rho_{0}$ due to the detectable and undetectable errors, respectively. We will simply call them \emph{detectable and undetectable bias}. In the special case of $\vec{w} = \vec{I}$, i.e. the unmitigated case, $B_{u, \vec{I}}$ and $B_{d, \vec{I}}$ are simply the undetectable and detectable error probability, respectively. 
    
\begin{figure}[t]
        \centering
        \includegraphics[width = 0.45\textwidth]{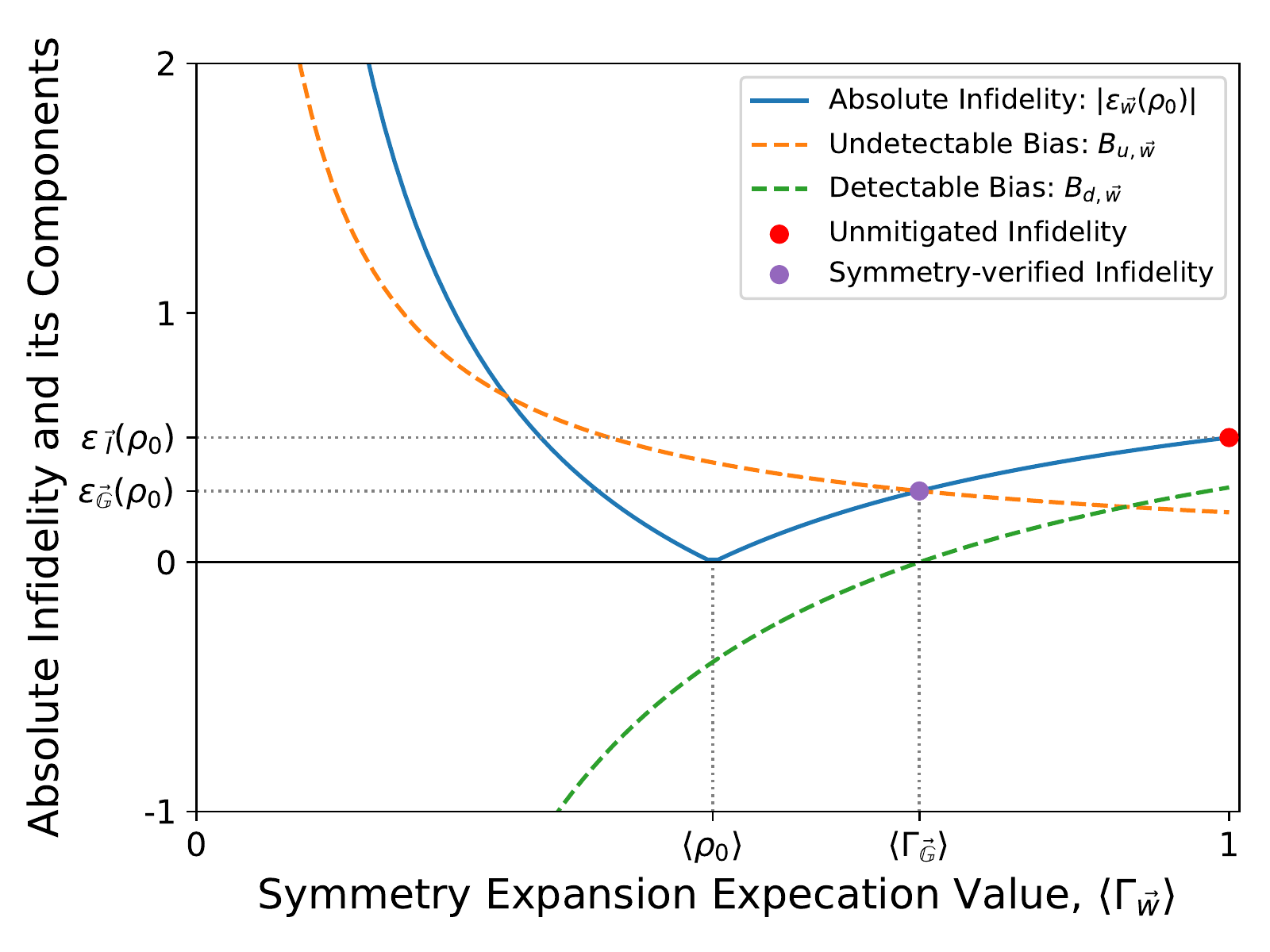}
        \caption{The change in the absolute infidelity along with its detectable and undetectable components against the change in the noisy expectation value of the symmetry expansion operator. We always have $\expval{\rho_0} \leq \expval{\Gamma_{\vec{\mathbb{G}}}}$ since the symmetry-verified fidelity in \cref{eqn:sym_fid} is always smaller than $1$. Note that the absolute infidelity is simply the magnitude of the sum of the detectable bias and the undetectable bias: $\abs{\epsilon_{\vec{w}}(\rho_0)} = \abs{B_{u, \vec{w}} + B_{d, \vec{w}}}$. }
        \label{fig:abs_expand_err}
    \end{figure}

    From \cref{eqn:abs_infid}, we can see that symmetry expansion with negative $\expval{\Gamma_{\vec{w}}}$ will lead to $\abs{\epsilon_{\vec{w}}(\rho_0)} \geq 1 \geq \abs{\epsilon_{\vec{I}}(\rho_0)}$, i.e. the corresponding symmetry expansion scheme will performs worse than the unmitigated case, thus there is no point of applying symmetry expansion in this regime. Therefore, we will only consider symmetry expansion operators with positive expectation values: $ 0 < \expval{\Gamma_{\vec{w}}} \leq 1$. In this regime, $B_{u, \vec{w}}$ will increase as $\expval{\Gamma_{\vec{w}}}$ decrease and will always be positive,  while $B_{d, \vec{w}}$ will decrease as $\expval{\Gamma_{\vec{w}}}$ decrease and will become negative once $\expval{\Gamma_{\vec{w}}} < \expval{\Gamma_{\vec{\mathbb{G}}}}$.
    Hence, the expression of the absolute infidelity can be split into the following regions:
    \begin{align*}
        \abs{\epsilon_{\vec{w}}(\rho_0)} = \begin{cases}
            B_{u, \vec{w}} + B_{d, \vec{w}}& 1 \geq \expval{\Gamma_{\vec{w}}} \geq \expval{\Gamma_{\vec{\mathbb{G}}}}\\
            B_{u, \vec{w}} - \abs{B_{d, \vec{w}}}& \expval{\Gamma_{\vec{\mathbb{G}}}} > \expval{\Gamma_{\vec{w}}} \geq \expval{\rho_0}\\
            \abs{B_{d, \vec{w}}} - B_{u, \vec{w}} & \expval{\rho_0} > \expval{\Gamma_{\vec{w}}} > 0
        \end{cases}
    \end{align*}
    which is plotted in \cref{fig:abs_expand_err}. Let us  now go through these different regions of symmetry expansion.
    
    \subsubsection{Standard behaviour: $1 \geq \expval{\Gamma_{\vec{w}}} \geq \expval{\Gamma_{\vec{\mathbb{G}}}}$}
    
    Here both the detectable bias $B_{d, \vec{w}}$ and the undetectable bias $B_{u, \vec{w}}$ are positive and thus the absolute infidelity is simply
    \begin{align*}
        \abs{\epsilon_{\vec{w}}(\rho_0)} = B_{u, \vec{w}} + B_{d, \vec{w}}.
    \end{align*}
    
    At the start of this region we have $\expval{\Gamma_{\vec{w}}} = \expval{I} = 1$, which is simply the noisy unmitigated case. As we modify our symmetry expansion scheme to decrease $\expval{\Gamma_{\vec{w}}}$, the undetectable bias $B_{u, \vec{w}}$ will increase while the detectable bias $B_{d, \vec{w}}$ will decrease. In this region, the absolute infidelity $\abs{\epsilon_{\vec{w}}(\rho_0)}$ will decrease overall as a result of decrease in $\expval{\Gamma_{\vec{w}}}$.
    
    When we reach the point of $\expval{\Gamma_{\vec{w}}} = \expval{\Gamma_{\vec{\mathbb{G}}}}$, we have arrived at symmetry verification, in which we have $B_{d, \vec{\mathbb{G}}} = 0$. Hence, symmetry verification have removed all the detectable error components and we have $\abs{\epsilon_{\vec{w}}(\rho_0)} = B_{u, \vec{w}}$.
    
    \subsubsection{Bias cancellation: $\expval{\Gamma_{\vec{\mathbb{G}}}} > \expval{\Gamma_{\vec{w}}} \geq \expval{\rho_0}$}
    In this region the detectable bias $B_{d, \vec{w}}$ turns to negative and thus the overall error becomes
    \begin{align*}
        \abs{\epsilon_{\vec{w}}(\rho_0)} = B_{u, \vec{w}} - \abs{B_{d, \vec{w}}} 
    \end{align*}
    We see that since the detectable bias $B_{d, \vec{w}}$ and the undetectable bias $B_{u, \vec{w}}$ have different signs, the absolute infidelity become the \emph{difference} between the magnitude of detectable and undetectable biases, allowing \emph{cancellation} between the detectable and undetectable bias and thus reducing the overall bias. 

    In this region, the absolute infidelity $\abs{\epsilon_{\vec{w}}(\rho_0)}$ will still decrease overall as a result of decrease in $\expval{\Gamma_{\vec{w}}}$. As we arrive at the point of $\expval{\Gamma_{\vec{w}}} = \expval{\rho_0}$, we have $ B_{u, \vec{w}} = \abs{B_{d, \vec{w}}}$, which means perfect cancellation between the detectable and undetectable bias, and thus the absolute infidelity is reduce to $\abs{\epsilon_{\vec{w}}(\rho_0)} = 0$.

    \subsubsection{Negative infidelity: $0 \leq \expval{\Gamma_{\vec{w}}} \leq \expval{\rho_0}$}
    In this region, we still have negative detectable bias $B_{d, \vec{w}}$, and thus there will still be cancellation between the detectable and undetectable biases. However, now the detectable bias has grown to a larger magnitude than the undetectable bias and thus in this region we have $\epsilon_{\vec{w}}(\rho_0)$ being negative, which means:
    \begin{align*}
        \abs{\epsilon_{\vec{w}}(\rho_0)} = \abs{B_{d, \vec{w}}}  - B_{u, \vec{w}}. 
    \end{align*}
    The absolute infidelity $\abs{\epsilon_{\vec{w}}(\rho_0)}$ will increase as we decrease $\expval{\Gamma_{\vec{w}}}$. When $\expval{\Gamma_{\vec{w}}}$ decrease beyond $\left(2 \expval{\rho_0}^{-1} - \expval{\Gamma_{\vec{\mathbb{G}}}}^{-1}\right)^{-1}$, we will have $\abs{\epsilon_{\vec{w}}(\rho_0)} > \abs{\epsilon_{\vec{\mathbb{G}}}(\rho_0)}$, i.e. symmetry expansion will have a larger bias than symmetry verification. When $\expval{\Gamma_{\vec{w}}}$ decrease beyond $\left(2 \expval{\rho_0}^{-1} - 1\right)^{-1}$, we will have $\abs{\epsilon_{\vec{w}}(\rho_0)} > \abs{\epsilon_{\vec{I}}(\rho_0)}$, i.e. symmetry expansion will have a larger bias than the unmitigated state. 
    
    \subsubsection{Simple example of bias cancellation}
    As explained above, some symmetry expansion schemes can achieve smaller absolute infidelity than symmetry verification due to the cancellation between the detectable and undetectable bias. We will further illustrate this through the simple example in which there is just one single non-trivial symmetry operator: $\mathbb{G} = \{I, G\}$. 
    
    We will compare the performance of three schemes:
    \begin{itemize}
        \item Unmitigated $\vec{w} = \vec{I}$: $\expval{\Gamma_{\vec{w}}} = 1$
        \item Symmetry-verified $\vec{w} = \vec{\mathbb{G}}$: $\expval{\Gamma_{\vec{w}}} = \expval{\Gamma_{\vec{\mathbb{G}}}} = \frac{1 + \expval{G}}{2}$
        \item Symmetry-expanded $\vec{w} = \vec{G}$: $\expval{\Gamma_{\vec{w}}} = \expval{G}$
    \end{itemize}
    
    As mentioned, the undetectable/detectable biases for the unmitigated case are simply the probability that some errors undetectable/detectable by the symmetry $G$ have occurred in a given circuit run, which we will denote as $P_u$/$P_d$:
    \begin{align*}
        P_u &= B_{u, \vec{I}} = \expval{\Gamma_{\vec{\mathbb{G}}}} - \expval{\rho_0}\\
        P_d &= B_{d, \vec{I}} = 1 - \expval{\Gamma_{\vec{\mathbb{G}}}} 
    \end{align*}
    Hence, we can write the absolute infidelities of various schemes in terms of $P_u$ and $P_d$ using \cref{eqn:abs_infid}:
    \begin{equation}
        \begin{aligned}\label{eqn:sym_err_prob}
            \text{Unmitigated: }&\epsilon_{\vec{I}}(\rho_0) =  P_u + P_d\\
            \text{Symmetry-verified: }&\epsilon_{\vec{\mathbb{G}}}(\rho_0) =  \frac{P_u }{1-P_d}\\
            \text{Symmetry-expanded: }&\abs{\epsilon_{\vec{G}}(\rho_0)} =  \abs{\frac{P_u - P_d}{1 - 2P_d} }.
        \end{aligned}
    \end{equation}
    We see that $\epsilon_{\vec{I}}(\rho_0)$ is proportional to the \emph{sum} of $P_u$ and $P_d$ since none of the errors are removed. We also have $\epsilon_{\vec{\mathbb{G}}}(\rho_0)$ being proportional to just $P_u$ since the detectable errors are removed by symmetry verification. On the other hand, $\abs{\epsilon_{\vec{G}}(\rho_0)}$ is proportional to the \emph{difference} between $P_u$ and $P_d$, allowing \emph{cancellation} between detectable and undetectable error probability, which is why symmetry expanding with $\vec{w} = \vec{G}$ can achieve a smaller bias than symmetry verification.

    \section{Implementation Costs of Symmetry Expansion}\label{sec:implementation_cost}
     \subsection{Implementation}\label{sec:pauli_implementation}
    
    In \emph{direct} symmetry verification, we need to measure the whole set of symmetry generators and the observable in the same circuit run, and discard runs that fail \emph{any} of the symmetry tests. By using additional circuit structures to simultaneously diagonalise all the observables and symmetries we want to measure~\cite{jenaPauliPartitioningRespect2019, hugginsEfficientNoiseResilient2021}, we would only need to perform local measurements directly on the data register without needing any ancilla. The additional circuit structure required can be vastly simplified (or even removed) by using \emph{post-processing} symmetry verification and more generally symmetry expansion, for which we only need to measure \emph{one} symmetry operator along with the observables in each circuit run. The Fermi-Hubbard simulation that we will study in our numerical simulation later in \cref{sec:simulation} is an example of no additional circuit structure required to perform our measurements. The detailed measurement scheme for this case can be found in Ref.~\cite{caiResourceEstimationQuantum2020}. 
    
    Since we are discarding circuit runs that fail the symmetry test in direct symmetry verification, we are essentially changing the form of the effective error channels in the circuit. Therefore combining direct symmetry verification with other error mitigation techniques like quasi-probability and error extrapolation would require elaborate schemes~\cite{caiMultiexponentialErrorExtrapolation2021} since they all rely on our knowledge about the error channels. On the other hand, there is no discarding circuit runs in symmetry expansion, and thus it can be combined with other error mitigation techniques straightforwardly,
    e.g. by simply performing symmetry expansion following \cref{eqn:exp_obs} using  $\expval{O\Gamma_{\vec{w}}}$ and $\expval{\Gamma_{\vec{w}}}$ that are error-mitigated using the other error mitigation techniques. 
    
    \subsection{Sampling Cost}\label{sec:sampling_cost}
    Now looking beyond the circuit implementation, we would want to know how many circuit runs are needed for symmetry expansion. For a given error mitigation scheme we have defined the corresponding sampling cost in \cref{eqn:sampling_cost} as the factor of increase in the number of circuit runs required compared to the unmitigated case. As shown in \cref{sec:sym_cost}, the sampling cost for implementing \emph{direct} symmetry verification and symmetry expansion are
    \begin{equation}
        \begin{aligned}\label{eqn:cost_qem}
            C_{dir} &\sim \expval{\Gamma_{\vec{\mathbb{G}}}}^{-1}, \quad
            C_{\vec{w}} &\sim \expval{\Gamma_{\vec{w}}}^{-2}.
        \end{aligned}
    \end{equation}
    As can be seen from \cref{sec:good_sym_cond}, in the region where $\expval{\Gamma_{\vec{w}}} \geq \expval{\rho_0}$ (i.e. $\epsilon_{\vec{w}}(\rho_{0})$ is positive), the symmetry-expanded fidelity behaves like regular fidelity and thus can be use as a metric for the estimation bias of symmetry expansion. For all symmetry expansions in this region, including post-processing symmetry verification, if we are allowed $C$ sampling cost for our error mitigation technique, symmetry expansion can give us a factor of $\sqrt{C}$ boost in the fidelity based on \cref{eqn:exp_fid} and \cref{eqn:cost_qem}. Hence, we see that symmetry expansion in this regime is as `cost-effective' as post-processing symmetry verification. Direct symmetry verification, on the other hand, can achieve a factor of $C$ boost in the fidelity and thus is more `cost-effective' but is more difficult to implement due to the measurement requirement and harder to combine with other error mitigation techniques as discussed in \cref{sec:pauli_implementation}.
    
    In the region where $\expval{\Gamma_{\vec{w}}} < \expval{\rho_0}$, the symmetry-expanded fidelity is not well-defined anymore and thus our metric for estimation bias becomes the absolute infidelity as discussed in \cref{sec:good_sym_cond}. In this region we can achieve the same bias as the symmetry expansion in the $\expval{\Gamma_{\vec{w}}} \geq \expval{\rho_0}$ region, but at a smaller $\expval{\Gamma_{\vec{w}}}$ and thus at a larger sampling cost $C_{\vec{w}}$. Hence, symmetry expansion in this region is less `cost effective' than post-processing symmetry verification.
    
    \section{Searching for Suitable Symmetry Expansion Schemes}\label{sec:find_opt_sym}
    \subsection{Estimation of Fidelity and Symmetry Expectation Value}\label{sec:exp_est}
    As discussed in \cref{sec:sym_exp_bias} and \cref{sec:implementation_cost}, we can estimate the bias and the sampling cost of a given symmetry expansion scheme using the expectation value of its symmetry expansion operator $\expval{\Gamma_{\vec{w}}}$ and the unmitigated fidelity $\expval{\rho_0}$. Hence, in order to compare among different symmetry expansion schemes, we need to first perform estimations of $\expval{\rho_0}$ and $\expval{\Gamma_{\vec{w}}}$. 
    
    We will call the expected number of errors occurring in each circuit run the \emph{mean circuit error count} and denote it using $\mu$. In practical NISQ applications, we would need to have a circuit that is large enough such that the number of error locations is much greater than $1$ and is not too noisy such that the mean circuit error count is of order unity: $\mu \sim 1$. In this limit, using the Le Cam's theorem, the probability that $n$ errors occurring in a given circuit run will follow the Poisson distribution provided the noise is Markovian: 
    \begin{align*}
        P_n =e^{-\mu}\frac{\mu^n}{n!}.
    \end{align*} 
    Hence, the probability that there is no errors in the circuit will be:
    \begin{align*}
        P_0 = e^{-\mu}.
    \end{align*}
    Let us use $f_\epsilon$ to denote the effective fraction of errors that would move the ideal state $\ket{\psi_0}$ out of the 1D subspace it spanned. Then the average number of \emph{effective} errors occurring on the ideal state is simply $f_{\epsilon} \mu$, and the fidelity $\expval{\rho_0}$ is just the probability that zero effective errors happens:
    \begin{align}\label{eqn:noi_fid}
        \expval{\rho_0} \sim P_0 = e^{-f_{\epsilon}\mu}.
    \end{align}
    In practice, we usually assume $f_{\epsilon} = 1$ for crude estimation of $\expval{\rho_0}$, which gives $\expval{\rho_0}  = e^{-\mu}$.

    Moving on, using \cref{eqn:exp_op}, $\expval{\Gamma_{\vec{w}}}$ is simply
    \begin{align}\label{eqn:gamma_exp}
        \expval{\Gamma_{\vec{w}}} = \frac{\sum_{G \in \mathbb{G}} w_G \expval{G}}{\sum_{G \in \mathbb{G}} w_G}.
    \end{align}
    Hence, we can estimate $\expval{\Gamma_{\vec{w}}}$ by simply measuring $G$ with sampling weight $w_G$ at the end of the circuit. 
    
    \subsection{Search for a Small-bias Symmetry Expansion} \label{sec:near_opt_scheme}
    In many of the practical application, due to the large amount of circuit noise present, very often the main bottleneck of quantum error mitigation is still trying to achieve a small enough estimation bias. Therefore, we hope to use symmetry expansion to further reduce the estimation bias of symmetry-based error mitigation method beyond symmetry verification. Here we will discuss how to search for a symmetry expansion scheme that has a smaller absolute infidelity and thus a smaller bias than symmetry verification. 
    
    The absolute infidelity $\abs{\epsilon_{\vec{w}}(\rho_0)}$ of a given symmetry expansion can be estimated using $\expval{\rho_0}$ and $\expval{\Gamma_{\vec{w}}}$ using \cref{eqn:abs_infid}. However, due to the approximation we made in estimating $\expval{\rho_0}$ and $\expval{\Gamma_{\vec{w}}}$, it is usually hard to fine-tune our expansion weights $\vec{w}$ such that the absolute infidelity $\abs{\epsilon_{\vec{w}}(\rho_0)}$ reach the exact value we want. On the other hand, the precision of our $\abs{\epsilon_{\vec{w}}(\rho_0)}$ estimates is often enough for comparing among the restricted class of symmetry expansion scheme mentioned in \cref{sec:sym_expand}, in which we only consider a \emph{uniform} expansion using a subset of symmetry operators $\mathbb{F} \subseteq \mathbb{G}$. For such symmetry expansion schemes, we have:
    \begin{align*}
        \expval{\Gamma_{\vec{\mathbb{F}}}} = \frac{1}{\abs{\mathbb{F}}}\sum_{F \in \mathbb{F}} \expval{F}.
    \end{align*}

    In order for the symmetry expansion scheme with $\vec{w} = \vec{\mathbb{F}}$ to achieve an absolute infidelity $\abs{\epsilon_{\vec{\mathbb{F}}}(\rho_0)}$ smaller than some threshold $\delta$, using \cref{eqn:abs_infid} we need to have:
    \begin{align}\label{eqn:sym_exp_threshold}
        \frac{\expval{\rho_{0}}}{1+\delta} \leq \expval{\Gamma_{\vec{\mathbb{F}}}} \leq \frac{\expval{\rho_{0}}}{1-\delta}.
    \end{align}
    Finding symmetry expansion schemes that has a smaller bias than symmetry verification is simply the case of $\delta = \epsilon_{\vec{\mathbb{G}}}(\rho_0)$.
    
    One way to construct symmetry expansions that fulfil this requirement is by first identifying the set of symmetry operator $\mathbb{R} \subseteq \mathbb{G}$ consists of all the symmetry operators whose noisy expectation value falling within this range:
    \begin{align}\label{eqn:sym_op_threshold}
        \frac{\expval{\rho_{0}}}{1+\delta} \leq \expval{R} \leq \frac{\expval{\rho_{0}}}{1-\delta} \quad \forall R \in \mathbb{R}.
    \end{align}
    In such a way, any subset $\mathbb{F} \subseteq \mathbb{R}$ will always satisfy \cref{eqn:sym_exp_threshold}. 
    
    Using \cref{eqn:sym_exp_threshold} we see that the scheme that has a smaller $\abs{\expval{\Gamma_{\vec{\mathbb{F}}}} - \expval{\rho_0}}$ can achieve a smaller estimation bias. Following \cref{eqn:noi_fid} and assuming $f_\epsilon \approx 1$, we have $\expval{\rho_0} \approx e^{-\mu}$. Hence, we should be able to achieve a small estimation bias using the set of symmetry operators that minimise $\abs{\expval{\Gamma_{\vec{\mathbb{F}}}} - e^{-\mu}}$, which is denoted as
    \begin{align}\label{eqn:opt_expand}
        \mathbb{G}^* = \argmin_{\mathbb{F} \subseteq \mathbb{R}} \abs{\expval{\Gamma_{\vec{\mathbb{F}}}} - e^{-\mu}}.
    \end{align}
    We need to be careful about the case in which we have two schemes whose $\expval{\Gamma_{\vec{\mathbb{F}}}}$ are of similar distance to $e^{-\mu}$, one from above and one from below. In this case, the scheme with larger $\expval{\Gamma_{\vec{\mathbb{F}}}}$ is expected to lead to a smaller bias as it will have $\expval{\Gamma_{\vec{\mathbb{F}}}}$ closer to the actual $\expval{\rho_0}$ (which is lower-bounded by $e^{-\mu}$), and it is also expected to have a smaller sampling cost as discussed in \cref{sec:sampling_cost}. Hence, the scheme with larger $\expval{\Gamma_{\vec{\mathbb{F}}}}$ is generally preferred in such a case. We will call the symmetry expansion scheme found in the way outlined above the \emph{small-bias expansion scheme}. 
    
    In \cref{sec:abs_inf_detail} and later in the numerical simulation (\cref{sec:simulation}), we can see that we usually have $\epsilon_{\vec{w}}(O) \sim \epsilon_{\vec{w}}(\rho_0)$. Hence,  using \cref{eqn:Nc_relative}, the minimum number of samples required for the small-bias expansion scheme ($\vec{w} = \vec{\mathbb{G}}^*$) to outperform symmetry verification ($\vec{w} = \vec{\mathbb{G}}$) for a Pauli observable is simply:
    \begin{align}\label{eqn:n_thresh_exp_ver}
        N^* \lesssim  \frac{C_{\vec{\mathbb{G}}^*} -C_{\vec{\mathbb{G}}}}{\epsilon_{\vec{\mathbb{G}}}(\rho_0)^2 - \epsilon_{\vec{\mathbb{G}}^*}(\rho_0)^2}.
    \end{align}

    \subsection{Alternative Search Strategy using Knowledge of Noise}\label{sec:search_using_f}
    When we have detailed knowledge about the noise channels, we can try to estimate 
    $\expval{\Gamma_{\vec{w}}}$ analytically instead. For a given Pauli symmetry $G$, if we can approximately decomposed all noise channels within each error location into components that are detectable by $G$ and components that are undetectable, with $f_G$ being the fraction of errors detectable by $G$ averaged over all error locations, then as shown in \cref{sec:sym_expec_est}, we have:
    \begin{align}\label{eqn:sym_est}
        \expval{G} \sim e^{-2f_G \mu}.
    \end{align} 
    Note that $f_I = 0$. Using \cref{eqn:gamma_exp}, we then have:
    \begin{align}\label{eqn:gamma_est}
        \expval{\Gamma_{\vec{w}}}  \sim \frac{\sum_{G \in \mathbb{G}} w_G e^{-2f_{G} \mu }}{\sum_{G \in \mathbb{G}} w_G}.
    \end{align}

    We can use \cref{eqn:noi_fid} and \cref{eqn:sym_est} to rewrite \cref{eqn:sym_op_threshold} at the circuit error rate $\mu$ as:
    \begin{align*}
        \frac{e^{-f_{\epsilon}\mu}}{1+\delta} \leq e^{-2f_R \mu} \leq \frac{e^{-f_{\epsilon}\mu}}{1-\delta} \quad \forall R \in \mathbb{R}.
    \end{align*}
    In the limit of small $\delta$, the first order approximation of the inequality above gives:
    \begin{align*}
         - \delta \leq -\left(2f_R - f_{\epsilon}\right) \mu \leq   \delta \quad \forall R \in \mathbb{R}.
    \end{align*}
    i.e.
    \begin{align*}
        2f_R \mu = f_{\epsilon}\mu + \order{\delta} \quad \forall R \in \mathbb{R}.
    \end{align*}
    In this small $\delta$ limit, the relative bias of symmetry expanding using any $\mathbb{F} \subseteq \mathbb{R}$ can be obtain using \cref{eqn:abs_infid}:
    \begin{align}\label{eqn:small_bias}
        \abs{\epsilon_{\vec{\mathbb{F}}}(\rho_0)} &= \abs{1 - \frac{\abs{\mathbb{F}}e^{-f_{\epsilon}\mu}}{\sum_{F \in \mathbb{F}} e^{-2f_F\mu}}} \approx \frac{1}{\abs{\mathbb{F}}}\abs{\sum_{F \in \mathbb{F}} \left(2 f_F - f_{\epsilon}\right)}\mu,
    \end{align}
    which is of $\order{\delta}$ as expected. The corresponding sampling cost can be obtained using \cref{eqn:cost_qem}:
    \begin{align}\label{eqn:small_bias_cost}
        C_{\vec{\mathbb{F}}} &= \frac{\abs{\mathbb{F}}^2}{\left(\sum_{F \in \mathbb{F}} e^{-2f_F\mu}\right)^2} \sim e^{-2f_{\epsilon}\mu}.
    \end{align}
    We see that the sampling cost grows exponentially with the mean circuit error count $\mu$, similar to other mainstream quantum error mitigation techniques~\cite{endoHybridQuantumClassicalAlgorithms2021}.
    
    Using \cref{eqn:small_bias} with $f_{\epsilon} \approx 1$, the search for the small-bias expansion scheme in \cref{eqn:opt_expand} can now be written as
    \begin{align}\label{eqn:opt_expand_f}
        \mathbb{G}^* = \argmin_{\mathbb{F} \subseteq \mathbb{R}} \frac{1}{\abs{\mathbb{F}}}\abs{\sum_{F \in \mathbb{F}} \left(2 f_F - 1\right)}. 
    \end{align}
    Similarly to our discussion before, when we have two schemes with similar $\frac{1}{\abs{\mathbb{F}}}\abs{\sum_{F \in \mathbb{F}} \left(2 f_F - 1\right)}$, the scheme with smaller $\frac{1}{\abs{\mathbb{F}}}\sum_{F \in \mathbb{F}} f_F $ will have larger $\expval{\Gamma_{\vec{w}}}$ and thus is preferred.

    \subsection{Reducing Search Space using Equivalent Symmetry}\label{sec:reduce_search_space}
    If a given symmetry operator $S$ commutes with the noisy state $\rho$, which also implies $S^{-1}$ commuting with $\rho$, we then have:
    \begin{align}\label{eqn:equiv_expec}
        \expval{G} = \Tr(S^{-1}SG\rho) = \Tr(SGS^{-1}\rho) = \expval{SGS^{-1}}.
    \end{align}
    One possible way that such a symmetry $S$ can arise is when it corresponds to some qubit permutations. In such a case, we would naturally design a state preparation circuit with a layout that respects this symmetry. Since we usually assume the nature of the noise of a given type of circuit components is the same for all of its instances in the circuit, we can expect the resultant noisy state $\rho$ coming out of this circuit will also follow the same symmetry: $S\rho S^{-1} = \rho$, i.e. $\rho$ commutes with $S$.

    \cref{eqn:equiv_expec} implies we can obtain the same value for $\expval{\Gamma_{\vec{w}}}$ even if we sample $SGS^{-1}$ in place of $G$. Hence, $G$ and $SGS^{-1}$ are equivalent when we are considering $\expval{\Gamma_{\vec{w}}}$, which in turn means they are equivalent when trying to construct a symmetry expansion scheme with a given absolute infidelity and a given sampling cost.
    
    We will denote the set of symmetry operators that commute with $\rho$ as $\mathbb{S}$, which is a subgroup of $\mathbb{G}$. We say that two elements $G$ and $G'$ are equivalent if they give the same noisy expectation value. Alternatively, based on our arguments above, we can write this as
    \begin{align*}
        G \sim G' \quad \text{iff } \exists S \in \mathbb{S} \quad SGS^{-1} = G'.
    \end{align*}
    We can prove that this is an equivalence relation, and thus will partition the whole symmetry group into different equivalence classes. The symmetry operator within each equivalence class will have the same effect when sampled in symmetry expansion for calculating the fidelity against the ideal states. Hence, to find the suitable symmetry expansion, we only need to pick \emph{one element from each equivalence class} and sample over them instead of sampling over the whole symmetry group. This vastly reduces the search space for the suitable expansion weight distribution. In the extreme case that $\rho$ commutes with all symmetry operators $G \in \mathbb{G}$, then the equivalence classes are just the conjugacy classes of $\mathbb{G}$.
    
    \section{Numerical Simulation}\label{sec:simulation}
    
    \begin{figure*}[htbp]
        \centering
        \subfloat[8-Qubit]{\includegraphics[height = 0.32\textwidth]{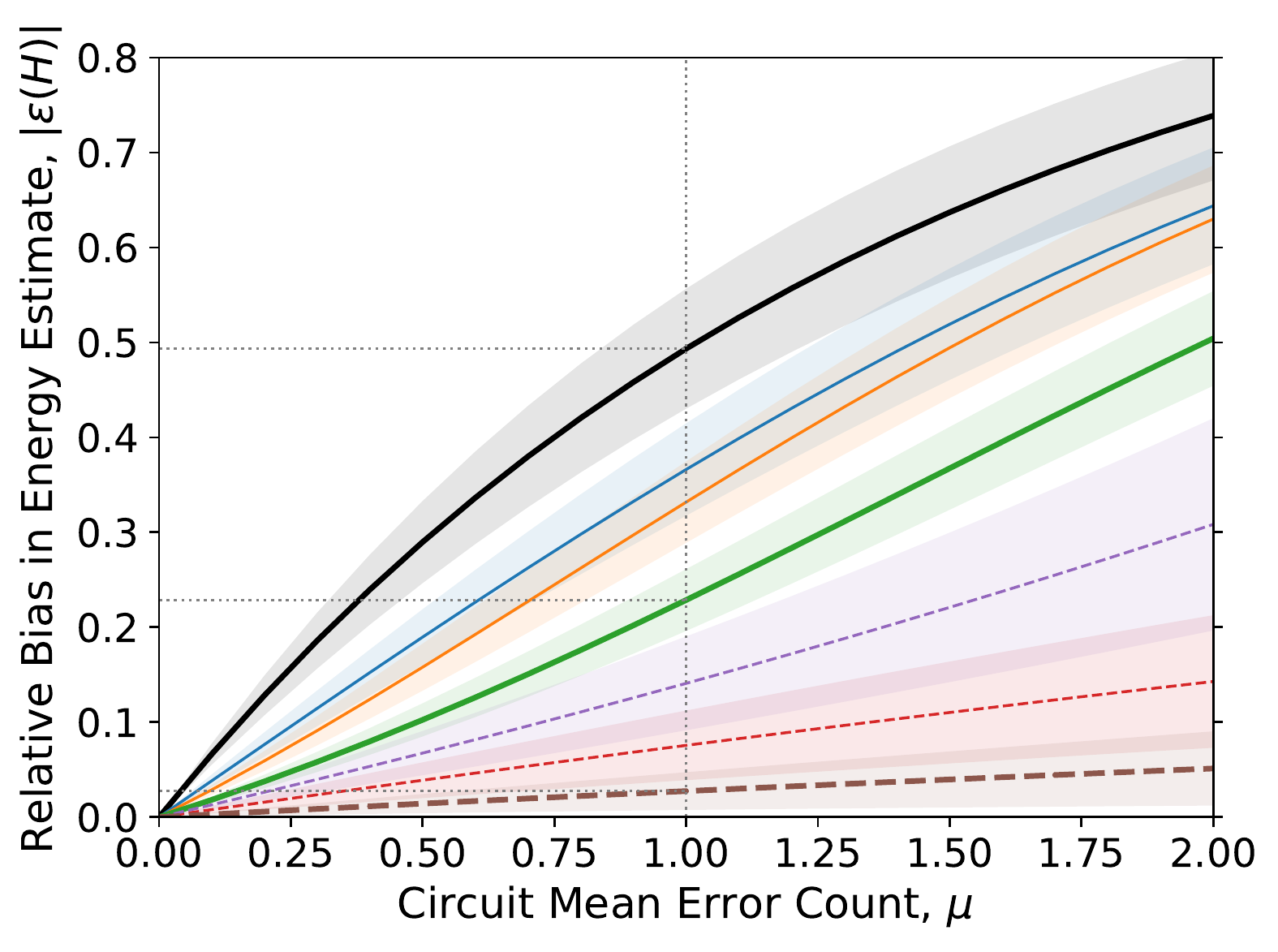}}\quad
        \subfloat[12-Qubit]{\includegraphics[height = 0.32\textwidth]{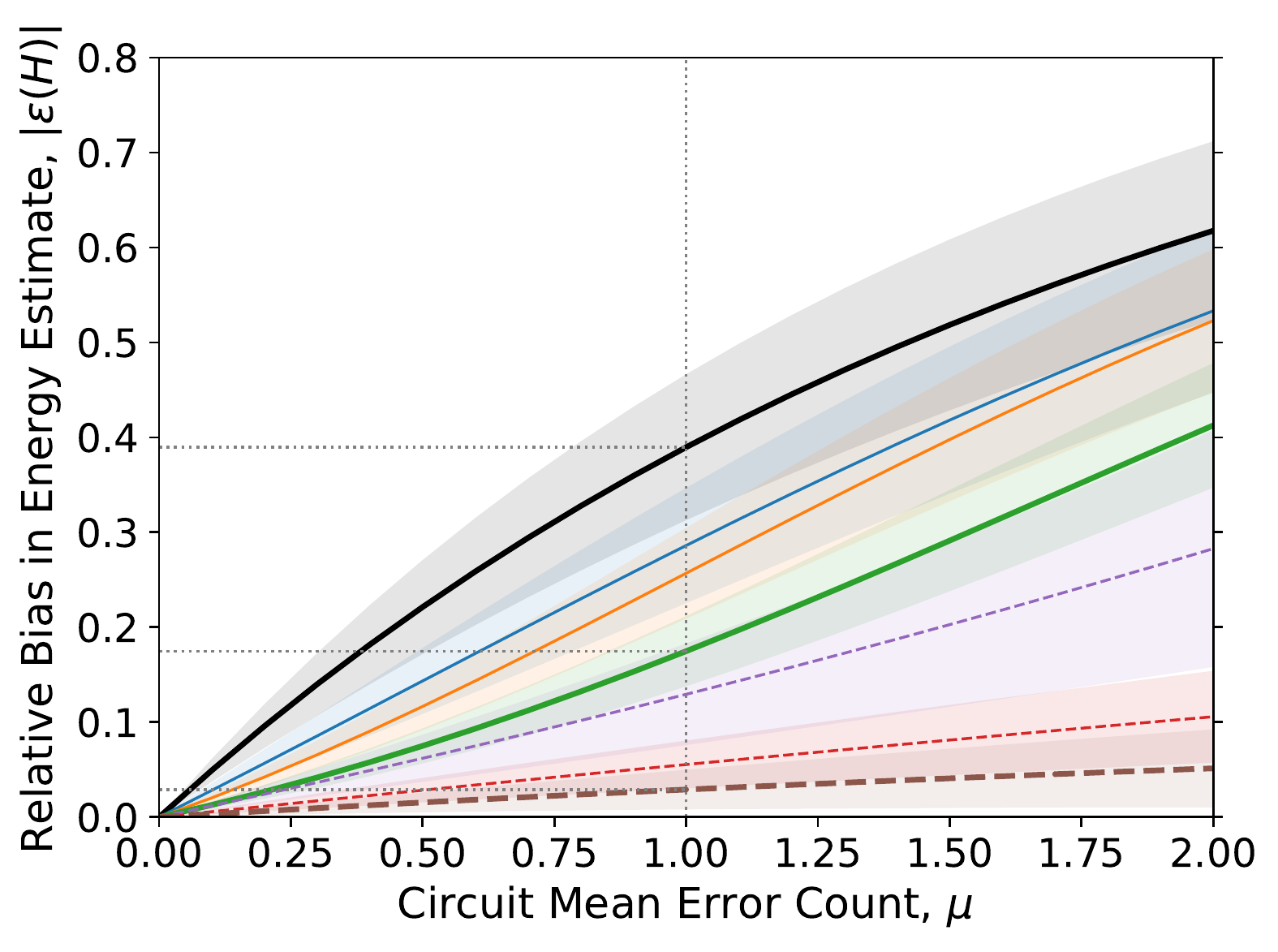}}\\
        \subfloat[8-Qubit]{\includegraphics[height = 0.32\textwidth]{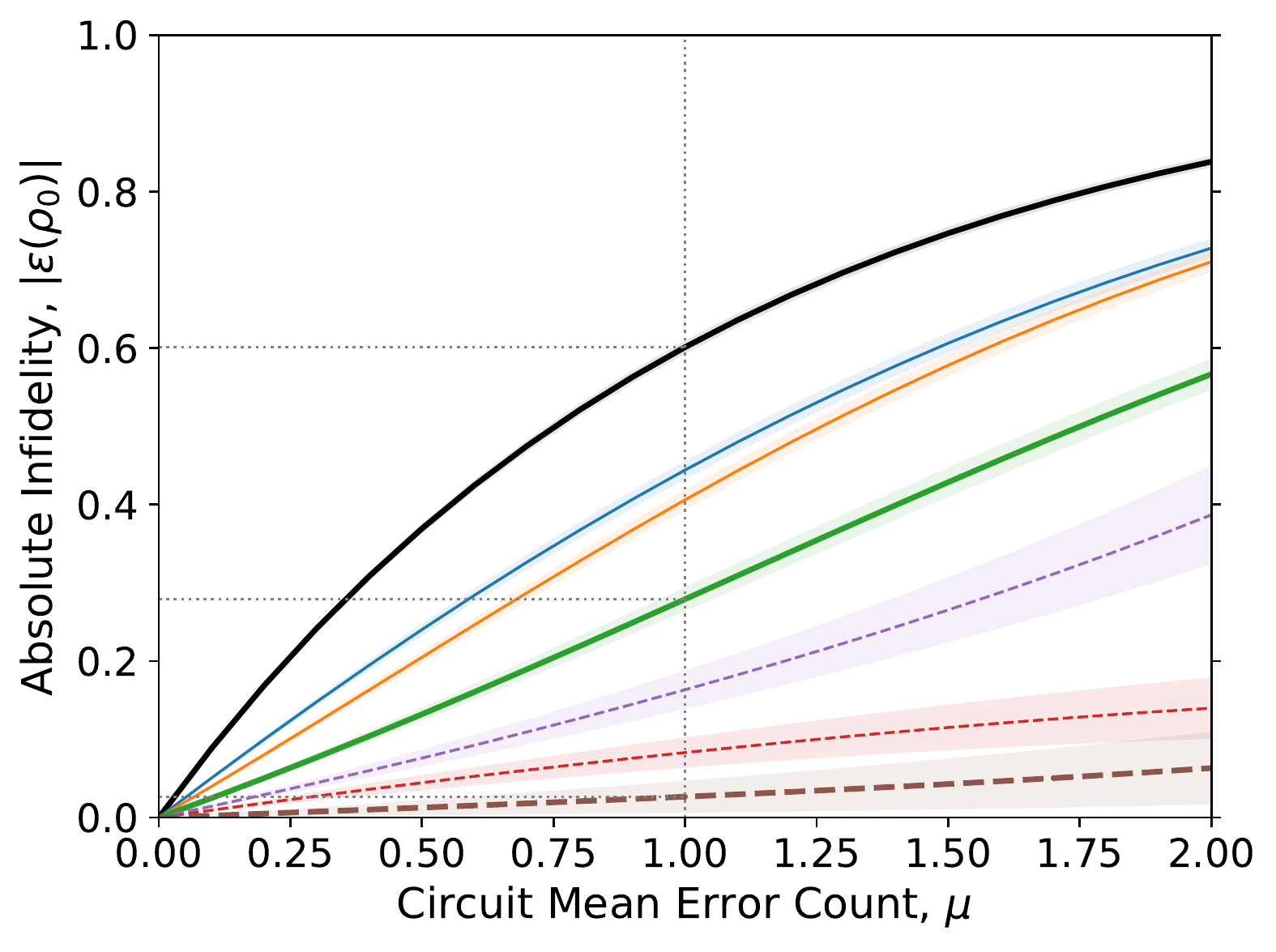}}\quad
        \subfloat[12-Qubit]{\includegraphics[height = 0.32\textwidth]{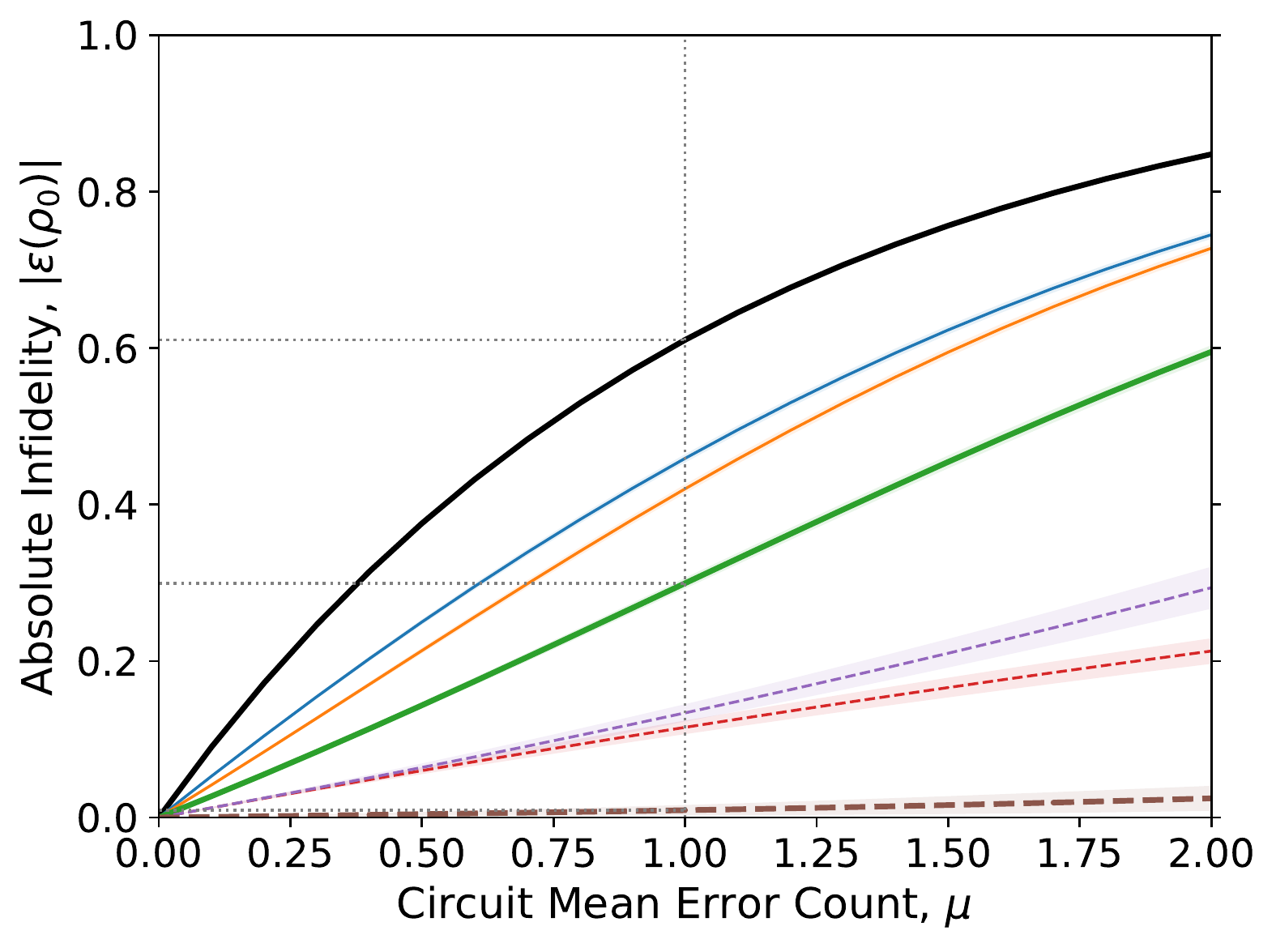}}\\
        \subfloat[]{\quad\quad\quad\quad \quad \includegraphics[height = 0.32\textwidth]{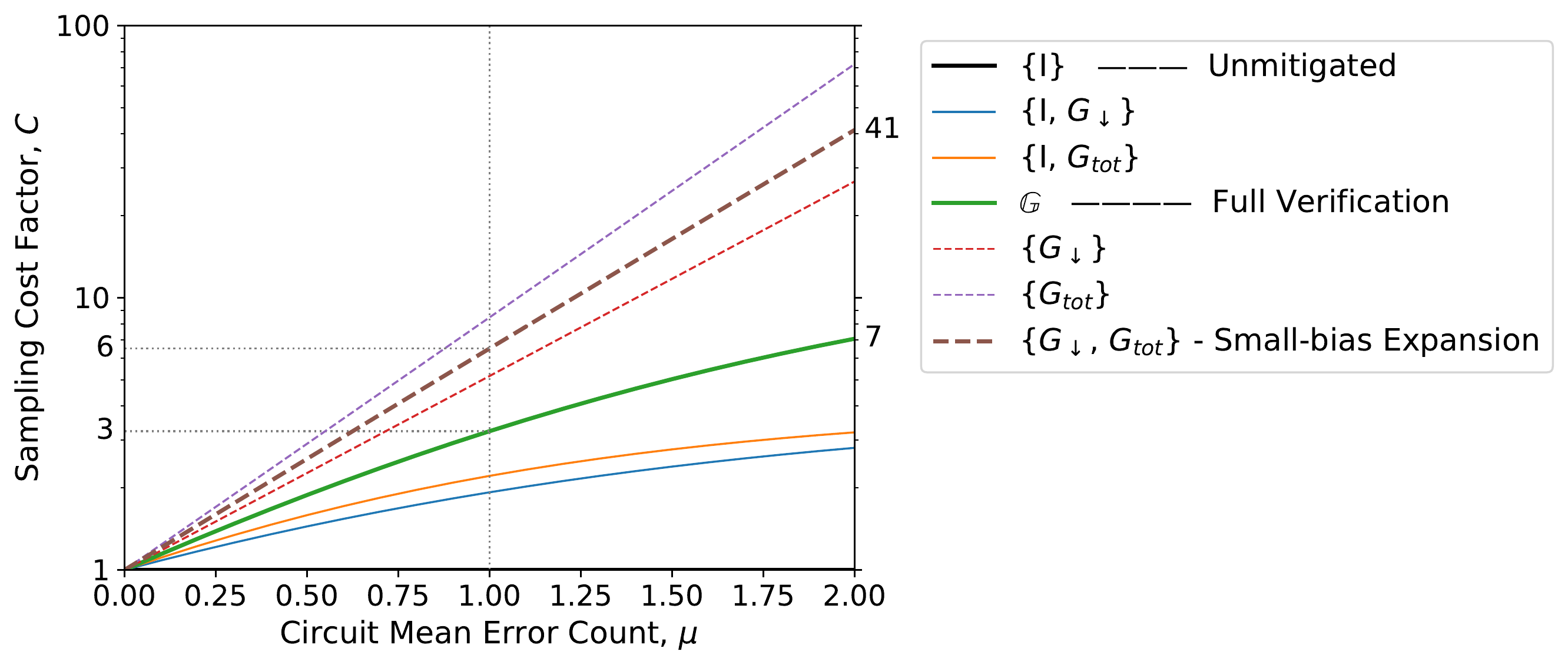}}
        \caption{Performance metrics for different symmetry expansion schemes against the increase of mean circuit error count under depolarising noise: the relative bias in the energy estimate for (a) 8-qubit and (b) 12-qubit simulations;  the absolute infidelity for (c) 8-qubit and (d) 12-qubit simulations; (e) the sampling cost for all simulations. Data are averaged over circuit configurations with randomly generated parameters. The shaded areas indicate the spread of the lines over different parameters. The symmetry expansions we consider are all uniform expansion of a given set of symmetry operators. The legend indicates the set of symmetry operators used in the corresponding symmetry expansion. When the set of symmetry operators forms a group, the corresponding symmetry expansion will be symmetry verification which is labelled using solid lines. All the other symmetry expansion are labelled using dashed lines. Three of the most representative methods: without mitigation (black), symmetry verification using the full symmetry group $\mathbb{G}= \{I,\ G_{\uparrow},\ G_{\downarrow},\ G_{tot}\}$ (green) and the small-bias symmetry expansion using  $\mathbb{G}^* = \{G_{\downarrow}, G_{tot}\}$ (brown)  are labelled with thicken lines.}
        \label{fig:sim_result}
    \end{figure*}
    
    \subsection{Problem Setup}
    The physical problem that we will be looking at is the 2D Fermi-Hubbard model with nearest-neighbour interaction, with the Hamiltonian given as:
    \begin{align} \label{eqn:hubbard_hamiltonian}
        \resizebox{\hsize}{!}{$
        H = \underbrace{-\sum_{\sigma, \expval{v, w}} t_{v, w} \left(a^\dagger_{v, \sigma} a_{w, \sigma} + a^\dagger_{w, \sigma} a_{v, \sigma}\right)}_{\text{nearest-neighbour hopping}} + \underbrace{\sum_v U_v n_{v, \uparrow}  n_{v, \downarrow}}_{\text{onsite repulsion}}.$}
    \end{align}
    Here $a^\dagger_{v, \sigma}/a_{v, \sigma}$ are the creation/annihilation operators of site $v$ with spin $\sigma$ and $n_{v, \uparrow/\downarrow}$ are the number operators. The exact circuit we used is outlined in Ref.~\cite{caiResourceEstimationQuantum2020}, which has the same form as a first-order trotterisation circuit. We use the Jordan-Wigner encoding to map the individual interaction terms and fermionic swaps to two-qubit gates in the circuit,  with the gates that correspond to the interaction terms parametrised. This circuit can be used in for example variational eigen-solvers. We would perform simulations for the $2\times 2$ and $2\times 3$ half-filled Fermi-Hubbard model, which corresponds to a 8-qubit circuit with $144$ two-qubit gates and a 12-qubit circuit with $336$ two-qubit gates, respectively. All two-qubit gates are assumed to be affected by two-qubit \emph{depolarising noise} of the same strength, which is of the form:
    \begin{align*}
        \mathcal{D}(\rho) = (1 - p) \rho + \frac{p}{\abs{\mathbb{P}_2}}\sum_{P \in \mathbb{P}_2} P\rho P
    \end{align*}
    where $\mathbb{P}_2$ is the full set of two-qubit Pauli operators. Similar simulations are also performed for bit-flip noise in \cref{sec:bit_flip_sim}, which yield similar results. 
    
    We will look at a range of circuits with randomly generated parameters and thus has random ideal output states. Using $t_{v, w} = 1$ and $U_v = 2$ for our Hamiltonian, we can measure the energy of the ideal output state and compare it to the energies of the noisy output state and the error-mitigated output states. In our simulation, we will focus on circuits whose ideal output energies \emph{magnitude} are larger than $0.5$ since in practice we are usually interested in states with non-negligible energy magnitude (otherwise the sampling cost for high energy precision will be very large). The simulation code used is available online~\cite{githubHttpsGithubCom2021}.

    \subsection{Different Symmetry Expansion Schemes}\label{sec:hubbard_sym}
    The ideal circuit will conserve the number of fermions within each spin subspace. Hence, the symmetry operator we can enforce on the output state would be the spin-up/down number parity symmetry, denoted as $G_{\uparrow/\downarrow}$. In the Jordan-Wigner encoding, $G_{\uparrow/\downarrow}$ is just the tensor product of the $Z$ operators acting on the qubits corresponding to the spin-up/down orbitals. Together they can generate the number parity symmetry group 
    \begin{align}\label{eqn:hubbard_full_sym}
        \mathbb{G} = \{I,\ G_{\uparrow},\ G_{\downarrow},\ G_{tot}\},
    \end{align}
    where $G_{tot} = G_{\uparrow}G_{\downarrow}$ is the total fermion number parity symmetry. As discussed in \cref{sec:near_opt_scheme}, our main focus will be on the symmetry expansion schemes that are uniform expansion of subsets of symmetry operators $\mathbb{F} \subseteq \mathbb{G}$. 
    
    The Hamiltonian in \cref{eqn:hubbard_hamiltonian} is invariant under exchange of spin up and spin down. This spin-exchange symmetry is denoted as $S$, and will naturally lead to a circuit construction that respects $S$.  Hence, as shown in \cref{sec:reduce_search_space}, we can prove that $G_{\uparrow}$ and the corresponding conjugated symmetry $SG_{\uparrow}S^{-1} = G_{\downarrow}$ would be equivalent when used for symmetry expansion. Note that for our practical implementation of the circuit, the trotter step implemented actually does not \emph{exactly} commute with $S$, thus
    the performance of $G_{\uparrow}$ and $G_{\downarrow}$ for symmetry expansion are only \emph{approximately} the same in our case. However, they are close enough such that the symmetry expansion schemes that have $G_{\downarrow}$ in place of $G_{\uparrow}$ or vice versa are considered to be identical to simplify our comparison between different expansions.
    
    As shown in \cref{sec:frac_of_detect_err}, the fraction of errors detectable by $G_{tot}$ is estimated to be $f_{G_{tot}} \sim \frac{8}{15}$ and that by $G_{\uparrow/\downarrow}$ is estimated to be $f_{G_{\uparrow/\downarrow}} \sim \frac{2}{5}$. Using these and following \cref{eqn:opt_expand_f} along the discussions there, the small-bias expansion scheme that we will find is simply uniform expansion with the set of symmetry operators:
    \begin{align}\label{eqn:hubbard_opt_exp}
        \mathbb{G}^* = \{G_{\uparrow/\downarrow}, G_{tot}\}.
    \end{align} 
    
    \subsection{Results}
    The absolute relative bias in the energy estimate, denoted as $\abs{\epsilon(H)}$ over different mean circuit error counts $\mu$ for the $8$-qubit and $12$-qubit simulations are shown in \cref{fig:sim_result} (a) and (b). The corresponding absolute infidelity $\abs{\epsilon(\rho_0)}$ are shown in \cref{fig:sim_result} (c) and (d). 
    From the close resemblance between the shapes of the curves for $\abs{\epsilon(H)}$ and $\abs{\epsilon(\rho_0)}$, we see that $\abs{\epsilon(\rho_0)}$ is indeed a good metric for comparing the estimation bias for different symmetry expansion methods as discussed in \cref{sec:sym_exp_bias}. This further validates our approach of searching for good symmetry expansions based on  $\abs{\epsilon(\rho_0)}$ in \cref{sec:find_opt_sym} and our assumption of $\epsilon_{\vec{w}}(O) \sim \epsilon_{\vec{w}}(\rho_0)$ in  \cref{eqn:n_thresh_exp_ver}.
    
    The symmetry expansion schemes that we look at are uniform expansions of different subsets of symmetry operators $\mathbb{F} \subseteq \mathbb{G}$ as discussed in \cref{sec:hubbard_sym}. When $\mathbb{F}$ forms a group, then symmetry expanding with $\mathbb{F}$ is simply performing symmetry verification using the symmetry group $\mathbb{F}$. We see that in our examples in \cref{fig:sim_result}, out of all symmetry verification schemes (solid lines), the symmetry verification using the full symmetry group $\mathbb{G}$ can achieve the lowest bias as expected. When we look at all symmetry expansion schemes including symmetry verifications, we see that the small-bias scheme we found in \cref{sec:hubbard_sym} can achieve the lowest bias. 
    
    The sampling costs are shown in \cref{fig:sim_result} (e), which is essentially the plot of the function $C_{\vec{w}} = \expval{\Gamma_{\vec{\mathbb{F}}}}^{-2}$ in \cref{eqn:cost_qem} for different $\mathbb{F}$. Since the estimated detectable error fraction $f_{G_{\uparrow/\downarrow}}$ and $f_{G_{tot}}$ are independent of the number of qubits and the circuit parameters, it means that $\expval{\Gamma_{\vec{w}}}$ is mostly independent of the number of qubits and the circuit parameters using \cref{eqn:gamma_est}, and thus our cost plot is practically the same for both the $8$-qubit and $12$-qubit cases and for circuits with any parameters. Note that the costs for the symmetry verifications here are for post-processing verification. If we go for direct verification instead, then we need to take the square root of the cost. 
    
    \begin{table}[t]
        \centering
        \begingroup
        \renewcommand{\arraystretch}{1.5}
        \setlength{\tabcolsep}{-1pt}
        \subfloat[$\mu=1$ ]{
            \begin{tabular}{lccccccc}\toprule
                & \multicolumn{2}{c}{8-Qubit} & \phantom{abc}& \multicolumn{2}{c}{12-Qubit}& \phantom{abc} & \multirow{2}{*}{Cost}\\
                \cline{2-3} \cline{5-6}
                & $\ \abs{\epsilon(H)}\ $ & $\ \abs{\epsilon(\rho_0)}\ $& &  $\ \abs{\epsilon(H)}\ $ & $\ \abs{\epsilon(\rho_0)}\ $ && \\ \hline
                Unmitigated & 0.493 & 0.601 &     & 0.390 & 0.610 &     &  1.0 \\
                Verified & 0.229 & 0.279 &     & 0.175 & 0.299 &     &  3.2 \\
                Expanded & 0.027 & 0.027 &     & 0.029 & 0.009 &     &  6.5 \\
                \botrule
        \end{tabular}}\\
        \subfloat[$\mu=2$ ]{
            \begin{tabular}{lccccccc}\toprule
                & \multicolumn{2}{c}{8-Qubit} & \phantom{abc}& \multicolumn{2}{c}{12-Qubit}& \phantom{abc} & \multirow{2}{*}{Cost}\\
                \cline{2-3} \cline{5-6}
                & $\ \abs{\epsilon(H)}\ $ & $\ \abs{\epsilon(\rho_0)}\ $& &  $\ \abs{\epsilon(H)}\ $ & $\ \abs{\epsilon(\rho_0)}\ $ && \\ \hline
                Unmitigated & 0.739 & 0.838 &     & 0.618 & 0.848 &     &  1.0 \\
                Verified & 0.504 & 0.567 &     & 0.413 & 0.595 &     &  7.1 \\
                Expanded & 0.051 & 0.063 &     & 0.051 & 0.025 &     & 41.4 \\
                \botrule
        \end{tabular}}
        \endgroup
        \caption{The relative bias in energy estimate $\abs{\epsilon(H)}$,  absolute infidelity $\abs{\epsilon(\rho_0)}$, and the sampling costs for three of the most representative symmetry expansions in our simulation: the unmitigated noisy state (Unmitigated), the symmetry verification using the full symmetry group $\mathbb{G}$ (Verified) and the small-bias symmetry expansion using  $\mathbb{G}^* = \{G_{\downarrow}, G_{tot}\}$ (Expanded) at the mean circuit error rate (a) $\mu = 1$ and (b) $\mu = 2$. Note that the cost for symmetry verification here are for post-processing symmetry verification. }
        \label{tab:est_err}
    \end{table}
    
    Let us zoom in at the mean circuit error count $\mu = 1$ and focusing on the three most representative symmetry expansions (thicken lines): the unmitigated noisy state, the symmetry verification using the full symmetry group $\mathbb{G}$ and the small-bias symmetry expansion using  $\mathbb{G}^* = \{G_{\downarrow}, G_{tot}\}$, whose data are summarised in \cref{tab:est_err} (a). We see that at $\mu = 1$, the energy estimate bias $\abs{\epsilon(H)}$ for the unmitigated noisy output sits at $\sim 0.5$ for the $8$-qubit example and  $\sim 0.4$ for the $12$-qubit example. Symmetry verification using $\mathbb{G}$ can reduce the bias to $\sim 0.2$, more than halving the bias, at the sampling cost of $\sim 3$. The bias in the energy estimate can be further reduced to $\sim 0.03$ when we apply the small-bias expansion scheme in \cref{eqn:hubbard_opt_exp}, which is $\sim 15$ times reduction compared to the unmitigated noisy state and $\sim 6$ times reduction compared to symmetry verification. The sampling cost for the symmetry expansion is $C_{\vec{\mathbb{G}}^*} \sim 6$, which is only twice the cost of symmetry verification. We see a similar, if not larger, improvement of the absolute infidelity $\abs{\epsilon(\rho_0)}$ by using the small-bias symmetry expansion scheme over the unmitigated noisy state and symmetry verification. Using \cref{eqn:n_thresh_exp_ver}, we have
    \begin{align*}
        N^* \lesssim \frac{C_{\vec{\mathbb{G}}^*} -C_{\vec{\mathbb{G}}}}{\epsilon_{\vec{\mathbb{G}}}(\rho_0)^2 - \epsilon_{\vec{\mathbb{G}}^*}(\rho_0)^2} \sim 40
    \end{align*}
    i.e. the minimum number of samples needed for the small-bias expansion to outperform symmetry verification for a Pauli observable is around $40$, which is a sampling number we can easily reach in practice and thus the small-bias scheme should be much preferred in practice.
    
    Now we will move to $\mu = 2$ with the data summarised in \cref{tab:est_err} (b). The bias in energy estimate $\abs{\epsilon(H)}$ for the unmitigated noisy output now rise to $\sim 0.74$ for the $8$-qubit example and $\sim 0.62$ for the $12$-qubit example. Symmetry verification can reduce the relative bias to $\sim 0.5$ and $\sim 0.4$ for the 8-qubit and 12-qubit case, respectively, with a sampling cost of $\sim 7$. The small-bias symmetry expansion on the other hand can still achieve a low relative bias of $\sim 0.05$, which is around $13$ times reduction compared to the unmitigated state and around $9$ times reduction compared to symmetry verification. Similar to the $\mu = 1$ case, again we see a similar if not larger improvements in the absolute infidelity $\abs{\epsilon(\rho_0)}$ compared to $\abs{\epsilon(H)}$ when we use the small-bias symmetry expansion. However, due to the high circuit error rate $\mu = 2$, the small-bias symmetry expansion would have a high sampling cost of $C_{\vec{\mathbb{G}}^*} \sim 41$. Using \cref{eqn:n_thresh_exp_ver}, we have
    \begin{align*}
        N^* \lesssim \frac{C_{\vec{\mathbb{G}}^*} -C_{\vec{\mathbb{G}}}}{\epsilon_{\vec{\mathbb{G}}}(\rho_0)^2 - \epsilon_{\vec{\mathbb{G}}^*}(\rho_0)^2} \sim 100
    \end{align*}
    i.e. the minimum number of samples needed for the small-bias expansion to outperform symmetry verification for a Pauli observable is around $100$. This is larger than the $\mu = 1$ case, but still reachable in practice. 
    
    Alternatively, at $\mu = 2$, we can opt for symmetry-expanding with only $G_{\downarrow}$, which can also achieve a much lower bias than symmetry verification at $\epsilon_{\vec{G}_{\downarrow}}(\rho_0) \approx 0.2$ as seen from \cref{fig:sim_result}, but at a lower sampling cost than the small-bias scheme at $C_{\vec{G}_{\downarrow}}\sim 27$. Hence, the minimum number of samples needed to outperform symmetry verification for a Pauli observable is around $ N^* \lesssim  70$ in this case, which is smaller than the small-bias scheme.
    
    \section{Conclusion}\label{sec:concl}
    In this article, we have constructed a general framework named \emph{symmetry expansion} for symmetry-based error mitigation techniques. Different symmetry expansions correspond to different sampling weight distributions among the symmetry operators, with symmetry verification corresponding to the uniform weight distribution over the full symmetry group. 
    The effective `density operator' after symmetry expansion is not positive semi-definite, thus instead of using the infidelity against the ideal state to predict the estimation bias of a given symmetry expansion, we introduced the metric \emph{absolute infidelity} against the ideal state to predict the estimation bias. Using the absolute infidelity as the metric, we have shown that some symmetry expansion schemes can achieve a smaller bias than symmetry verification through cancellation between the biases due to the detectable and undetectable noise components. For any given symmetry expansion scheme, we have shown ways to estimate its bias and sampling cost using the overall circuit error rate and measurements of the symmetry operators. Such performance prediction can be further simplified if we have knowledge about the error channels in the circuit. Using the bias prediction, we can search for the uniform symmetry expansion over some subset of symmetry operators that has the smallest predicted bias. The scheme we found is simply called the small-bias expansion scheme. 
    
    We then applied our methods to the energy estimation of Fermi-Hubbard model simulation with random circuit parameters for both the $8$-qubit and $12$-qubit cases. From our simulation, we see that indeed absolute infidelity is a good metric for predicting the bias in our energy estimate. When there is on average $1$ error per circuit run, the relative energy estimation biases can be reduced from $0.4 \sim  0.5$ for the unmitigated noisy state and  $\sim 0.2 $ for symmetry verification to $\sim 0.03$ for the small-bias symmetry expansion. The sampling cost for symmetry verification in this case is $3$, i.e. we need $3$ times more circuit runs for shot noise reduction, while the sampling cost for the small-bias symmetry expansion is around $6$. Hence, symmetry expansion can reach an estimation accuracy $6$ times beyond what is achievable by symmetry verification at a sampling cost that is only $2$ times higher. When there is on average $2$ errors per circuit run, we found that the factor of bias reduction brought by the small-bias symmetry expansion further increases, but the sampling cost also rise to $41$ due to the high circuit error rate. The small-bias scheme will still outperform symmetry verification given a reasonable amount of samples allowed, but we can also turn to another expansion schemes with a slightly higher bias (still lower than symmetry verification) at a lower sampling cost.
     
     In practice, choosing the error-mitigation scheme with the right bias-variance trade-off would largely depends on the precision we wish to reach and/or the number circuit runs allowed for our experiment. Symmetry expansion is able to provide us with a wider range of symmetry-based schemes beyond symmetry verification to fit our various practical requirements on the bias-variance trade-off. The estimation biases of the symmetry expansion can be further improved if we could develop effective ways to search for the relevant symmetry expansion schemes, which could be done by for example constructing a better metric beyond the absolute infidelity or using Clifford approximation learning-based method outlined in Ref.~\cite{strikisLearningbasedQuantumError2021}. 
    
    In this article, we have only considered positive sampling weights for symmetry expansion and it would be natural to extend this to negative sampling weights like in the quasi-probability error mitigation~\cite{temmeErrorMitigationShortDepth2017, endoPracticalQuantumError2018}. On the other hand, it would also be interesting to see if any method can be developed to modify the sampling distribution within quasi-probability like how we go from symmetry verification to symmetry expansion. Part of this has been explored in Ref.~\cite{strikisLearningbasedQuantumError2021}. 
    
    We have provided an explicit example of the application of symmetry expansion for Pauli symmetry group and for virtual distillation (\cref{sec:virt_distill}) in this article. One could look into the more detailed application of symmetry expansion in the other cases, e.g. non-abelian symmetry group. Note that non-abelian symmetry expansion is much easier to implement than direct non-abelian symmetry verification, as non-abelian symmetry verification cannot be done by simply measuring their generators. However in this case, the expectation values of the symmetry operators might not be real any more, and thus we need to redevelop our methods for finding a suitable symmetry expansion, possibly involving using complex weights. We might use methods in Ref.~\cite{mitaraiMethodologyReplacingIndirect2019} to measure such complex expectation values.
    
    All of our symmetry expansion arguments are directly applicable to the stabiliser symmetry group in the stabiliser codes. It would be interesting to look at an explicit example of how symmetry expansion might perform in the context of stabiliser code like what has been done for quantum subspace expansion~\cite{mccleanDecodingQuantumErrors2020}.  To find the optimal symmetry expansion in this context, one might want to draw inspiration from Ref.~\cite{caiMitigatingCoherentNoise2020} about ways to reduce the search space. In addition, it would also be interesting to see how symmetry expansion can be applied to the symmetry present in continuous variable systems like bosonic codes~\cite{albertSymmetriesConservedQuantities2014, grimsmoQuantumComputingRotationSymmetric2020, gertlerProtectingBosonicQubit2021}.

    \section*{Acknowledgements}
     The numerical simulations are performed using QuESTlink~\cite{jonesQuESTlinkMathematicaEmbiggened2020}, which is the Mathematica interface of the high-performance quantum computation simulation package QuEST~\cite{jonesQuESTHighPerformance2019}. The author is grateful to those who have contributed to both these valuable tools. 
    
    The author would like to thank Simon Benjamin and Balint Koczor for valuable discussions and Tyson Jones for his help on the usage of QuESTlink.
    
    The author is supported by the Junior Research Fellowship from St John’s College, Oxford and acknowledges support from the QCS Hub (EP/T001062/1).
    
    \appendix
    
    \section{Comparison between Error Mitigation Techniques}\label{sec:err_mit_comp}
Using $\overline{O}_{\mathrm{em}}$ to denote the sample mean of $O_{\mathrm{em}}$ after $N$ samples, the \emph{mean square error} of the error-mitigated sample mean estimator $\overline{O}_{\mathrm{em}}$ is simply:
\begin{align*}
    \resizebox{\hsize}{!}{$
    \mse{\overline{O}_{\mathrm{em}}} = \big\langle\left(\overline{O}_{\mathrm{em}} - O_0\right)^2\big\rangle
    = \bias{O_{\mathrm{em}}}^2 + \frac{\var{O_{\mathrm{em}}}}{N}.$}
\end{align*}
The root mean square error $\sqrt{\mse{\overline{O}_{\mathrm{em}}}}$ will indicate the estimation precision we can reach in practice given $N$ samples. We see that the contribution from the variance term can be reduced by increasing the number of samples $N$, but the overall precision is lower-bounded by the bias contribution $\abs{\bias{O_{\mathrm{em}}}}$.

Suppose we are required to obtain an answer whose root mean square error $\sqrt{\mse{\overline{O}_{\mathrm{em}}}}$ is below a certain error bound $\delta$, we need to first make sure our estimation bias is below the error bound:
\begin{align}\label{eqn:bias_cond}
    \abs{\bias{O_{\mathrm{em}}}} < \delta.
\end{align}

If both schemes have their biases below this threshold, then we will compare the number of samples required for them to reach this error threshold $\sqrt{\mse{\overline{O}_{\mathrm{em}}}} = \delta$:
\begin{align*}
    N_\delta = \frac{\var{O_{\mathrm{em}}}}{\delta^2 - \bias{O_{\mathrm{em}}}^2} 
\end{align*} 
The scheme with a smaller $N_\delta$ will be our go-to scheme. 

In this article, we do not have a fixed precision requirement for our estimators, thus we turn to another way to compare between different error mitigation schemes.

If we are given a fixed number of samples allowed instead of being given the required precision, and the two error mitigation schemes are labelled $1$ and $2$, one with a smaller bias $\bias{O_1}^2 < \bias{O_2}^2$ while the other with a smaller variance $\var{O_1} > \var{O_2}$, then the two schemes will achieve the same mean square error when the number of samples is:
\begin{align*}
    N^* = \frac{\var{O_1} - \var{O_2}}{\bias{O_2}^2 - \bias{O_1}^2}.
\end{align*}
When the number of allowed samples is more than this threshold $N > N^*$, then the error due to bias dominates and thus scheme $1$ with smaller bias is preferred, on the other hand when  $N < N^*$, then variance is the larger error source and thus scheme $2$ with smaller variance is preferred.

Using the expression of relative bias in \cref{eqn:relative_bias} and sampling cost in \cref{eqn:sampling_cost}, we have:
\begin{align*}
    N^* = \frac{C_1 - C_2}{\epsilon_{2}(O)^2 - \epsilon_{1}(O)^2} \frac{\var{O}}{\expval{O_0}^2}
\end{align*} 
For a Pauli observable $O$, we have $\var{O} = 1 - \expval{O}^2$. We usually only care about Pauli observable with expectation values $\expval{O}$ and $\expval{O_0}$ that have non-negligible magnitude, because otherwise a large number of samples is needed to reach high accuracy and the contribution of the Pauli observable to the actual physical properties we care about (which would be a linear sum of Pauli observables) is small. Hence, we have $\var{O} \lesssim \expval{O_0}^2 $ and thus
\begin{align}\label{eqn:Nc_relative_app}
    N^* \lesssim \frac{C_1 -C_2}{\epsilon_{2}(O)^2 - \epsilon_{1}(O)^2}.
\end{align}

\section{Identifying the Symmetry of the Output State}\label{sec:output_sym}
In a standard quantum circuit, we start with a quantum state $\ket{\psi_{init}}$ and perform a series of unitary operations (gates) on it by evolving it under different Hamiltonians, which will output a final state $\ket{\psi_{out}}$, using which we can perform measurements of some observables of interest $O$. To obtain an estimate of the expectation value of the observable $\bra{\psi_{out}}O\ket{\psi_{out}}$, we need to run the above circuit over and over again and take the average of the output.

When we try to prepare a certain state $\ket{\psi_{0}}$ for some physical problems, e.g. the ground state for a given Hamiltonian, we usually can identify some symmetries $G$ that stabilise the state:
\begin{align*}
    G \ket{\psi_{0}} = \ket{\psi_{0}}
\end{align*}
based on the known physical properties of the ideal state $\ket{\psi_{0}}$, e.g. the particle number, spin number, spatial symmetry, etc. 

To produce a final state $\ket{\psi_{out}}$ that stabilised by the right symmetry $G$ using our quantum circuit, we can start with an initial state is stabilised by the given symmetry:
\begin{align*}
    G\ket{\psi_{init}} = \ket{\psi_{init}}
\end{align*}
and restrict our circuit $U$ to those that commute with the symmetry
\begin{align*}
    \left[G, U\right] = 0.
\end{align*} 
Note that this usually means the circuit is built from blocks that commute with the symmetry. As a result,
when the circuit above is noiseless, the output state $\ket{\psi_{out}} = U\ket{\psi_{init}}$ will be stabilised by the given symmetry $G$:
\begin{align*}
    G\ket{\psi_{out}} = GU\ket{\psi_{init}} =  UG\ket{\psi_{init}} = \ket{\psi_{out}}.
\end{align*} 
In this case, any violation of the symmetry of the output state is due to the noise in the circuit, and performing symmetry verification will enable us to remove noise that leads to symmetry violation. 

On the other hand, if our circuit is not constructed in the way above and would output a noiseless state $\ket{\psi_{out}}$ that does not respect the symmetry $G$, then symmetry verification can project $\ket{\psi_{out}}$ into the same symmetry subspace as $\ket{\psi_{0}}$, which enables us to remove errors due to imperfect circuit representation for $\ket{\psi_{0}}$ besides the removal of noise in the circuit. 

In this article, we would assume that we have made a sensible choice of our state preparation circuit such that it produces $\ket{\psi_{out}}$ that respect the same set of symmetries as $\ket{\psi_{0}}$, as were most of the cases in practice, thus symmetry verification will remove the noise in the circuit as discussed above.

\section{Optimal Symmetry Subspace Expansion}\label{eqn:opt_sub_exp}
The subspace-expanded density operator with the expansion operator $\Gamma_{\vec{v}}$ is:
\begin{align}\label{eqn:subspace_exp}
    \rho_{\vec{v},s} = \frac{\Gamma_{\vec{v}} \rho \Gamma_{\vec{v}}^\dagger }{\expval{\Gamma_{\vec{v}}^\dagger \Gamma_{\vec{v}}}} = \frac{\sum_{i,j} v_jG_j \rho G_i^\dagger v_i^*  }{\sum_{i,j} v_i^*\expval{G_i^\dagger G_j}v_j }
\end{align}
and the corresponding expectation value for some observable $H$ is:
\begin{align}\label{eqn:energy_sub_exp}
    E_{\vec{v}} &=    \Tr(H\rho_{\vec{v},s}) =   \frac{\sum_{i,j} v_i^*\expval{G_i^\dagger H G_j}v_j }{\sum_{i,j} v_i^*\expval{G_i^\dagger G_j}v_j } = \frac{\vec{v}^\dagger \boldsymbol{H}\vec{v} }{\vec{v}^\dagger \boldsymbol{S} \vec{v} }
\end{align}
with
\begin{align}
    \boldsymbol{H}_{ij} = \expval{G_i^\dagger H G_j}\label{eqn:sub_H}\\
    \boldsymbol{S}_{ij} = \expval{G_i^\dagger G_j}.\label{eqn:sub_S}
\end{align}
Note that we are only interested in $\vec{v}$ that satisfy
\begin{align}\label{eqn:S_inner}
    \vec{v}^\dagger \boldsymbol{S} \vec{v} \neq 0
\end{align}
so that \cref{eqn:subspace_exp} and \cref{eqn:energy_sub_exp} are well-defined. 

Finding the extremal points in \cref{eqn:energy_sub_exp} is commonly done in the Rayleigh-Ritz method and is equivalent to solving the generalised eigenvalue problem:
\begin{align}\label{eqn:gen_eigen}
    \boldsymbol{H} \vec{v} = E_{\vec{v}} \boldsymbol{S}\vec{v}.
\end{align}

In the case of $H = \rho_0 = \ket{\psi_0} \bra{\psi_0}$ for some ideal \emph{pure} state $\ket{\psi_0}$, we have $E_{\vec{v}}$ being the fidelity of the subspace-expanded state against the ideal state following \cref{eqn:energy_sub_exp}. Since all $G$ are symmetry of $\rho_0$ we have:
\begin{align*}
    \boldsymbol{H}_{ij} = \expval{G_k^\dagger \rho_0 G_i} = \expval{ \rho_0 } \quad \forall i,j
\end{align*}
Hence, $\boldsymbol{H}$ is a matrix with all entries being $\expval{ \rho_0 }$. This is a rank $1$ matrix that has only one non-zero eigenvalue $\abs{\mathbb{G}}\expval{ \rho_0 }$ whose corresponding eigenvector is $\vec{1}$.

Since the set of $G_i$ forms a group $\mathbb{G}$ , we see that the rows of $\boldsymbol{S}$ are simply permutations of the set of expectation value $\{\expval{G}\ |\ G \in \mathbb{G}\}$, and thus $\vec{1}$ will be an eigenvector of $\boldsymbol{S}$ with the eigenvalue $\sum_{G \in \mathbb{G}} \expval{G}$.

Hence, we see that $\vec{v} = \vec{1}$  would satisfy \cref{eqn:gen_eigen} with $E_{\vec{v}}$ given by:
\begin{align*}
    \boldsymbol{H} \vec{1} &= E_{\vec{v}} \boldsymbol{S}\vec{1}\\
    \abs{\mathbb{G}}\expval{ \rho_0 } \vec{1} &= E_{\vec{v}} \sum_{G \in \mathbb{G}} \expval{G} \vec{1}\\
    E_{\vec{v}} &= \frac{\abs{\mathbb{G}}\expval{ \rho_0 }}{\sum_{G \in \mathbb{G}} \expval{G}} = \frac{\expval{ \rho_0 }}{\expval{\Gamma_{\vec{\mathbb{G}}}}}
\end{align*}
For any other eigenvectors of $\boldsymbol{S}$, denoted as $\vec{v}_{\perp}$, we will only be interested in those with non-zero eigenvalue $s_{\vec{v}_{\perp}}$ due to \cref{eqn:S_inner}. Since $\vec{v}_{\perp}$ is orthogonal to $\vec{1}$, we have $\boldsymbol{H} \vec{v}_{\perp} = 0$. Hence, $\vec{v} = \vec{v}_{\perp}$ would satisfy \cref{eqn:gen_eigen} with $E_{\vec{v}}$ given by:
\begin{align*}
    0 &= E_{\vec{v}} \boldsymbol{S}\vec{v}_{\perp}\\
    0 &= E_{\vec{v}} s_{\vec{v}_{\perp}} \vec{v}_{\perp}\\
    E_{\vec{v}} &= 0
\end{align*}
i.e. all extremal points other than $\vec{v} = \vec{1}$ would given zero fidelity $E_{\vec{v}} = 0$.

Hence, the symmetry subspace expansion weights that would maximise the fidelity when all our symmetry operators form a full group is always $\vec{v} = \vec{1}$, i.e. it is simply performing symmetry verification using the symmetry group.

\section{Linking Infidelity to Relative Bias}\label{sec:abs_inf_detail}
We can extract out the noiseless component of the symmetry-expanded state $\rho_{\vec{w}}$ by writing it as:
\begin{align}\label{eqn:noi_decomp}
    \rho_{\vec{w}} &= F_{\vec{w}} \rho_0 +  \left(1 - F_{\vec{w}}\right) \rho_{\varepsilon, \vec{w}}.
\end{align}
where $F_{\vec{w}} = \Tr(\rho_{\vec{w}}\rho_0)$ is the coefficient for the error-free component as expected, and erroneous component $\rho_{\varepsilon, \vec{w}}$ is simply
\begin{align*}
    \rho_{\varepsilon, \vec{w}} = \frac{(1-\rho_0) \rho_{\vec{w}}}{\Tr((1-\rho_0) \rho_{\vec{w}})} = \frac{(1-\rho_0) \rho_{\vec{w}}}{1 - F_{\vec{w}}}.
\end{align*}
which has unit trace but is not necessarily positive semi-definite. By definition we have $\Tr(\rho_0\rho_{\varepsilon,\vec{w}}) = 0$.

Using \cref{eqn:frac_err_def} and \cref{eqn:noi_decomp}, the relative bias for expectation value estimation using symmetry expansion can be written as 
{\small
\begin{equation}
    \begin{aligned}
        \epsilon_{\vec{w}}(O) &= 1 - \frac{\Tr(O\rho_{\vec{w}})}{\Tr(O \rho_{0})} = 1 - \left(F_{\vec{w}} + (1 - F_{\vec{w}}) \frac{\Tr(O\rho_{\varepsilon, \vec{w}})}{\Tr(O \rho_{0})}\right)\\
        &= \epsilon_{\vec{w}}(\rho_0) \epsilon_{\varepsilon, \vec{w}}(O)
    \end{aligned}
\end{equation}
}
where $\epsilon_{\vec{w}}(\rho_0)  = 1 - F_{\vec{w}}$ is the infidelity of the symmetry expansion and
\begin{align*}
    \epsilon_{\varepsilon, \vec{w}}(O) &= 1 - \frac{\Tr(O\rho_{\varepsilon, \vec{w}})}{\Tr(O\rho_0)}
\end{align*}
is the relative bias of the erroneous component $\rho_{\varepsilon, \vec{w}}$.

In practice, we are usually only interested in the traceless part of the observable since the trace only corresponds to a uniform shift of the spectrum of the observable. Hence, without loss of generality we will assume our observable of interests $O$ is traceless. Now in practice, the ideal output state $\rho_0$ is usually dominated by the eigenstates of the observable of interest $O$ that lies close to the extreme of the spectrum (e.g. the ground state) such that the magnitude of $\Tr(O\rho_0)$ is not too small to be measured to high accuracy. The noisy component $\rho_{\varepsilon, \vec{w}}$ on the other hand is usually not dominated by any eigenstates of $O$ as it is the result of random noise. In such a case, we usually have $\Tr(O\rho_{\varepsilon, \vec{w}})) \ll \Tr(O\rho_0)$ and thus the relative bias of the erroneous component would be close to $1$: $\epsilon_{\varepsilon, \vec{w}}(O) \sim 1$, which means:
\begin{equation}
    \begin{aligned}
        \epsilon_{\vec{w}}(O) &\sim \epsilon_{\vec{w}}(\rho_0).
    \end{aligned}
\end{equation}
In \cref{sec:simulation}, from numerical simulations we see that even if some of the assumptions above are violated, e.g. when $\rho_0$ does not lie close to extreme of the spectrum of $O$, $\epsilon_{\vec{w}}(\rho_0)$ still serve as a good metric for $\epsilon_{\vec{w}}(O)$.

\section{Estimation of Symmetry Expectation Values}\label{sec:sym_expec_est}
Arguments in this section largely follows from Ref.~\cite{caiMultiexponentialErrorExtrapolation2021}. We will assume the average number of errors in each circuit run is $\mu$. Now looking within each error location in the circuit, there will be only a fraction of the errors that can be detected by a given symmetry $G$. We will denote this detectable error fraction \emph{averaged over all error locations} as $f_G$. Correspondingly, it means that the average number of such detectable errors in each circuit run will be $f_G \mu$. If the number of error locations with non-negligible detectable error components is much larger than $1$, then following the same arguments in \cref{eqn:exp_fid} while focusing on only detectable errors, the probability that $n$ \emph{detectable} errors occur in a given circuit run is:
\begin{align*}
    P_{n} = e^{-f_G\mu}\frac{\left(f_G\mu\right)^n}{n!}.
\end{align*}
For a Pauli symmetry $G$, only an odd number of such detectable errors will be detected and an even number of such detectable errors will pass the symmetry test. Hence, using $\Pi_{G}$ to denote the projector into the $G = +1$ subspace, the probability of passing the symmetry test using $G$ is just:
\begin{align*}
    \expval{\Pi_{G}} =  \sum_{\text{even } n} P_n = e^{-f_G\mu} \cosh(f_G\mu) = \frac{1 + e^{-2f_G\mu}}{2}.
\end{align*}
Hence, we have: 
\begin{align*}
    \expval{G} = 2\expval{\Pi_{G}} -1 = e^{-2f_G\mu}.
\end{align*}

Hence, the more general projection operator with multiple Pauli symmetries is simply:
\begin{align}\label{eqn:general_pi_est}
    \expval{\Pi_{\mathbb{G}}} \approx \frac{1}{\abs{\mathbb{G}}}\sum_{G \in \mathbb{G}} e^{-2f_{G}\mu}.
\end{align}
where $f_G$ is the fraction of errors that are detectable by $G$. Note that $f_{I} = 0$. 

Now we look at a group of Pauli symmetry $\mathbb{G}$ generated by the generators $\widetilde{\mathbb{G}}$. If the errors detectable by different symmetry generators $\widetilde{G} \in \widetilde{\mathbb{G}}$ are independent, i.e. each error component is \emph{detectable and only detectable} by one of the symmetry generators $\widetilde{G} \in \widetilde{\mathbb{G}}$, then the symmetry tests using different symmetry generators will be independent. Hence, we have:
\begin{align*}
    \expval{\Pi_{\mathbb{G}}} \approx \prod_{\widetilde{G} \in \widetilde{\mathbb{G}}}\expval{\Pi_{\widetilde{G}}} = \prod_{\widetilde{G} \in \widetilde{\mathbb{G}}}\frac{1 + e^{-2f_{\widetilde{G}}\mu}}{2}.
\end{align*}
which is a special case of \cref{eqn:general_pi_est}.

Here we have implicitly assumed that the errors in the circuit are \emph{incoherent} mixture of detectable errors and undetectable errors such that the average number of detectable errors occurring in each circuit run $f_G \mu$ is well-defined. If this is not true, then we can always try to decohere the detectable and undetectable errors via \emph{twirling}~\cite{wallmanNoiseTailoringScalable2016}.

\section{Variance of Quotient}
If we have a random variable $C$ that is obtained by taking the quotient of two other random variables $A$ and $B$:
\begin{align*}
    C =\frac{A}{B},
\end{align*}
then the variance of $C$ can be approximated using:
\begin{align}\label{eqn:var_quotient_sing}
    \resizebox{\hsize}{!}{$
    \var{C} = \frac{1}{\expval{B}^2} \left[\var{A} - 2\expval{C} \cov{A,B} + \expval{C}^2 \var{B}\right].$}
\end{align}
Throughout this section, we will use $\overline{X}$ to denote the estimator of $\expval{X}$ after $N$ \emph{circuit runs}. If in each circuit run, we can obtain one sample of $A$ and one sample of $B$, then by definition, we have:
\begin{align*}
    \overline{C} = \frac{\overline{A}}{\overline{B}}
\end{align*}
and
\begin{equation}
    \begin{aligned}\label{eqn:var_quotient}
        &\quad \var{\overline{C}} \\
        &= \frac{1}{\expval{B}^2} \left[\var{\overline{A}} - 2\expval{C} \cov{\overline{A},\overline{B}} + \expval{C}^2 \var{\overline{B}}\right]\\
        &= \frac{1}{N\expval{B}^2} \left[\var{A} - 2\expval{C} \cov{A,B} + \expval{C}^2 \var{B}\right].
    \end{aligned}
\end{equation}

\section{Sampling Cost Analysis}\label{sec:sym_cost}
In this section, we will follow the discussion in \cref{sec:sym_expand}, in which our circuit outputs the noisy quantum state $\rho$ and we want to estimate the value of a Pauli observable $O$ by using the symmetry verification with the symmetry projection operator $\Pi_{\mathbb{G}}$ or using symmetry expansion with the expansion operator $\Gamma_{\vec{w}}$. For a given observable $X$ whose noisy expectation value is  $\expval{X} = \Tr(X\rho)$, we will slightly abuse the notation and use $X$ to also denote the unbiased estimator of $\expval{X}$. 

\subsection{Direct Symmetry Verification}
The symmetry-verified expectation value of a Pauli observable $O$ is:
\begin{align}\label{eqn:osym}
    \expval{O_{dir}} = \frac{\expval{\Pi_{\mathbb{G}} O\Pi_{\mathbb{G}}}}{\expval{\Pi_{\mathbb{G}}}}.
\end{align}
Since $\Pi_{\mathbb{G}} O \Pi_{\mathbb{G}}$ takes the value 0, 1 and can be 1 if and only if $\Pi_{\mathbb{G}}$ takes the value 1, we have $\left(\Pi_{\mathbb{G}} O \Pi_{\mathbb{G}}\right)^2  = \Pi_{\mathbb{G}}^2 = \Pi_{\mathbb{G}}$. Using this and \cref{eqn:osym}, the variance and covariance of the observables are:
\begin{equation}
    \begin{aligned}\label{eqn:dir_comp_var}
        \var{\Pi_{\mathbb{G}} O \Pi_{\mathbb{G}}} &= \expval{\left(\Pi_{\mathbb{G}} O \Pi_{\mathbb{G}}\right)^2} - \expval{\Pi_{\mathbb{G}} O \Pi_{\mathbb{G}}}^2\\
        &=\expval{\Pi_{\mathbb{G}}} - \expval{O_{dir}}^2\expval{\Pi_{\mathbb{G}}}^2\\
        \var{\Pi_{\mathbb{G}}}  &= \expval{\Pi_{\mathbb{G}}^2} - \expval{\Pi_{\mathbb{G}}}^2 = \expval{\Pi_{\mathbb{G}}} - \expval{\Pi_{\mathbb{G}}}^2\\
        \cov{ \Pi_{\mathbb{G}} O \Pi_{\mathbb{G}}, \Pi_{\mathbb{G}}} &= \expval{\Pi_{\mathbb{G}} O \Pi_{\mathbb{G}}^2} - \expval{\Pi_{\mathbb{G}} O \Pi_{\mathbb{G}}}\expval{\Pi_{\mathbb{G}}}\\
        &= \expval{O_{dir}}\expval{\Pi_{\mathbb{G}}}\left(1- \expval{\Pi_{\mathbb{G}}}\right).
    \end{aligned}
\end{equation}
If we can measure all symmetry generators and $O$ in the same circuit run, then in each circuit run, we will obtain one sample for $\Pi_{\mathbb{G}} O\Pi_{\mathbb{G}}$ and one sample for $\Pi_{\mathbb{G}}$. Hence, after $N$ circuit runs, our estimation of the symmetry-verified variable is:
\begin{align}
    \overline{O}_{dir}=\frac{ \overline{\Pi_{\mathbb{G}} O \Pi_{\mathbb{G}}}}{\overline{\Pi_{\mathbb{G}}}}.
\end{align}
and the corresponding variance can be obtained using \cref{eqn:var_quotient} and \cref{eqn:dir_comp_var}:
\begin{equation}
    \begin{aligned}
        \label{eqn:dir_var}
        &\quad \var{\overline{O}_{dir}} \\
        &= \frac{1}{N\expval{\Pi_{\mathbb{G}}}} \bigg[\left(1 - \expval{O_{dir}}^2\expval{\Pi_{\mathbb{G}}}\right) -  \expval{O_{dir}}^2\left(1- \expval{\Pi_{\mathbb{G}}}\right)\bigg]\\
        &= \frac{1 -   \expval{O_{dir}}^2}{N\expval{\Pi_{\mathbb{G}}}} 
    \end{aligned}
\end{equation}

\subsection{Symmetry Expansion}
The symmetry-expanded expectation value of a Pauli observable $O$ is:
\begin{align}\label{eqn:osym_exp}
    \expval{O_{\vec{w}}} = \frac{\expval{O \Gamma_{\vec{w}}}}{\expval{\Gamma_{\vec{w}}}}.
\end{align}

We will decompose $O \Gamma_{\vec{w}}$ into:
\begin{align}
    O \Gamma_{\vec{w}} &= \frac{\sum_{G \in \mathbb{G}}  w_G OG}{\sum_{G' \in \mathbb{G}} w_{G'}}.
\end{align}
Hence, $O \Gamma_{\vec{w}}$ can be sampled by measuring different $OG$ with $\frac{w_G}{\sum_{G' \in \mathbb{G}} w_{G'}}$ probability. $\Gamma_{\vec{w}}$ can be sampled in a similar way by measuring $G$ instead of $OG$. To estimate $\expval{O_{\vec{w}}}$, in each circuit run we will measure the observable $O$ along with the symmetry operator $G$ with probability $\frac{w_G}{\sum_{G' \in \mathbb{G}} w_{G'}}$. They can be composed to obtain $OG$ and $G$, giving us one sample each to the variable $O \Gamma_{\vec{w}}$ and $\Gamma_{\vec{w}}$. Note that sampling in this way will means that $O \Gamma_{\vec{w}}$ and $\Gamma_{\vec{w}}$ are variables that return $\pm 1$ for each sample. Hence, we have $\left(O \Gamma_{\vec{w}}\right)^2 = \Gamma_{\vec{w}}^2 = 1$, and also $O \Gamma_{\vec{w}}\Gamma_{\vec{w}} = O \Gamma_{\vec{w}}^2 = O$. Using these and \cref{eqn:osym_exp}, the variance and covariance of the observables are:
\begin{equation}
    \begin{aligned}\label{eqn:exp_comp_var}
        \var{O \Gamma_{\vec{w}}} &= \expval{\left(O \Gamma_{\vec{w}}\right)^2} - \expval{O \Gamma_{\vec{w}}}^2 = 1 - \expval{O \Gamma_{\vec{w}}}^2 \\
        &= 1 - \expval{O_{\vec{w}}}^2\expval{\Gamma_{\vec{w}}}^2\\
        \var{\Gamma_{\vec{w}}} &= \expval{\Gamma_{\vec{w}}^2} - \expval{\Gamma_{\vec{w}}}^2  = 1 - \expval{\Gamma_{\vec{w}}}^2 \\
        \cov{ O \Gamma_{\vec{w}}, \Gamma_{\vec{w}}} &= \expval{O \Gamma_{\vec{w}} \Gamma_{\vec{w}}} - \expval{O \Gamma_{\vec{w}}}\expval{\Gamma_{\vec{w}}}\\
        &= \expval{O} -\expval{O_{\vec{w}}}\expval{\Gamma_{\vec{w}}}^2.
    \end{aligned}
\end{equation}
When we sample in the way outlined above, for each circuit run, we get one sample for $O \Gamma_{\vec{w}}$ and one sample for $\Gamma_{\vec{w}}$ just like in the direct verification case, it is just that these two variables are different random variable now. Hence, after $N$ circuit runs, our estimation of the symmetry-expanded variable is:
\begin{align}
    \overline{O}_{\vec{w}}=\frac{\overline{O \Gamma_{\vec{w}}}}{\overline{\Gamma_{\vec{w}}}}
\end{align}
and the corresponding variance can be obtained using \cref{eqn:var_quotient} and \cref{eqn:exp_comp_var}:
\begin{equation}
    \begin{aligned}\label{eqn:exp_var}
        &\quad \var{\overline{O}_{\vec{w}}} \\
        &= \frac{1}{N\expval{\Gamma_{\vec{w}}}^2} \bigg[1 - \expval{O_{\vec{w}}}^2\expval{\Gamma_{\vec{w}}}^2 \\
        &\quad - 2\expval{O_{\vec{w}}}\left(\expval{O} -\expval{O_{\vec{w}}}\expval{\Gamma_{\vec{w}}}^2\right) + \expval{O_{\vec{w}}}^2 \left(1 - \expval{\Gamma_{\vec{w}}}^2\right)\bigg] \nonumber\\
        &= \frac{1 - 2\expval{O_{\vec{w}}}\expval{O} + \expval{O_{\vec{w}}}^2}{N\expval{\Gamma_{\vec{w}}}^2}.
    \end{aligned}
\end{equation}

\subsection{Comparison}\label{sec:cost_comp}
When we directly sample from the noisy circuit, after $N$ circuit runs, the variance in the estimator of $O$ is:
\begin{align}\label{eqn:ideal_var}
    \var{\overline{O}} = \frac{\expval{O^2} - \expval{O}^2}{N} = \frac{1 - \expval{O}^2}{N},
\end{align}
which is upper-bounded by $\frac{1}{N}$. Compared to the upper-bounds of the variance using symmetry verification and symmetry expansion in \cref{eqn:dir_var} and \cref{eqn:exp_var}, the sampling cost factors are given by:
\begin{itemize}
    \item Direct Symmetry Verification: 
    \begin{align}
        C_{dir} = \frac{\var{\overline{O}_{dir}}}{\var{\overline{O}}} \sim \expval{\Pi_{\mathbb{G}}}^{-1} \equiv \expval{\Gamma_{\vec{\mathbb{G}}}}^{-1}
    \end{align}
    \item Symmetry Expansion (including post-processing symmetry verification):
    \begin{align}
        C_{\vec{w}} = \frac{\var{\overline{O}_{\vec{w}}}}{\var{\overline{O}}}  \sim \expval{\Gamma_{\vec{w}}}^{-2}.
    \end{align}
\end{itemize}

\section{Virtual Distillation}\label{sec:virt_distill}
Here we will revisit the virtual distillation protocol introduced in Refs.~\cite{koczorExponentialErrorSuppression2020, hugginsVirtualDistillationQuantum2021} from the perspective of symmetry expansion. Suppose we are interested in the ideal state $\ket{\phi_0}$. We can prepare $M$ copies of it: $\ket{\psi_0} = \ket{\phi_0}^{\otimes M}$, such that it is invariant under any permutation between the copies. Hence, its symmetry group is the permutation group of $M$ elements, denoting as $\mathbb{S}_M$. The projection operator into the subspace stabilised by $\mathbb{S}_M$ is simply:
\begin{align}\label{eqn:proj_vir_dist}
    \Pi_M = \frac{1}{\abs{\mathbb{S}_M}} \sum_{S \in \mathbb{S}_M} S =  \frac{1}{M!} \sum_{S \in \mathbb{S}_M} S.
\end{align}
In reality, we have $M$ copies of the noisy state $\rho = \sigma^{\otimes M}$ instead. Hence, we can implement symmetry verification using the permutation group to project the noisy state into the symmetry subspace defined by \cref{eqn:proj_vir_dist}~\cite{berthiaumeStabilisationQuantumComputations1994, barencoStabilizationQuantumComputations1997, peresErrorSymmetrizationQuantum1999}. 

Since the $M$ copies of the noisy state are identical, we know that $\rho = \sigma^{\otimes M}$ commute with all $S \in \mathbb{S}_M$, and thus commute with $\Pi_M$. Hence, the effective state we have after the symmetry verification is:
\begin{align}\label{eqn:sym_state_vir_dist}
    \rho_{\mathrm{sym}} = \frac{\Pi_M \rho \Pi_M }{\Tr(\Pi_M \rho \Pi_M )} &= \frac{\Pi_M \rho}{\Tr(\Pi_M \rho)}.
\end{align}
This is a special case of the symmetry expanded state in \cref{eqn:rho_exp} with $\mathbb{G} = \mathbb{S}_M$ and uniform weights. 

From here on, our discussion will deviate from the main text since in virtual distillation, we only care about the observable on \emph{the first copy}. If we observe an observable $O_1$ on the first copy of the symmetry-\emph{expanded} state, we have:
\begin{align*}
    \Tr(O_1\rho_{\vec{w}}) = \frac{\Tr(O_1\Gamma_{\vec{w}} \rho)}{\Tr(\Gamma_{\vec{w}} \rho )} = \frac{\sum_{S \in \mathbb{S}_M} w_S \Tr(O_1 S \rho)}{\sum_{S \in \mathbb{S}_M} w_S \Tr(S \rho )}.
\end{align*}

By direct calculation, we have
\begin{align*}
    \Tr(O_1S\rho) =  \Tr(O_1 S \sigma^{\otimes M}) = \Tr(O_1 \sigma^{m_S})
\end{align*}
where we use $m_S$ to denote the order of the cycle that copy $1$ belongs to in the cyclic representation of the permutation operator $S$. Hence, here using the structure of our problem, we have successfully identified sets of permutation operators within which all symmetry operators are equivalent for the purpose of being the components of symmetry expansion to observe $O_1$. Each set here contains permutation operators that have $1$ in a cycle of the same given order $m$. Let us define $C_{m}$ as the cyclic permutation operator acting on the \emph{first} $m$ copies, which is one of the operators in the set that has $1$ in an $m$-cycle, then we simply have:
\begin{align*}
    \Tr(O_1S\rho) = \Tr(O_1C_{m_S}\rho) \quad \forall S \in \mathbb{S}_M.
\end{align*}
Hence, our symmetry expansion operator only needs to have one element from each equivalence class, which are just $C_m$ of different $m$. It can be rewritten as:
\begin{align*}
    \Gamma_{\vec{w}} = \frac{\sum_{m = 1}^{M} w_mC_m}{\sum_{m = 1}^{M} w_m},
\end{align*}
and the symmetry projected observable is simply:
\begin{align*}
    \frac{\Tr(O_1  \Gamma_{\vec{w}} \rho )}{\Tr( \Gamma_{\vec{w}} \rho)} = \frac{\sum_{m = 1}^{M} w_m\Tr(O_1  \sigma^m )}{\sum_{m = 1}^{M} w_m\Tr(\sigma^m )}
\end{align*}
Now if $O_1$ is the ideal state $\sigma_0 = \ket{\phi_0} \bra{\phi_0}$, we then have fidelity of the symmetry expanded state against the ideal state of the first copy being:
\begin{align*}
    F = \frac{\Tr(\sigma_0  \Gamma_{\vec{w}} \rho )}{\Tr( \Gamma_{\vec{w}} \rho)} &= \frac{\sum_{m = 1}^{M} w_m\Tr(\sigma_0  \sigma^m )}{\sum_{m = 1}^{M} w_m\Tr(\sigma^m )}
\end{align*}
We can rewrite it as
\begin{align*}
    F_{\vec{p}}&= \sum_{m = 1}^{M} p_m   \frac{\Tr(\sigma_0  \sigma^m )}{\Tr(\sigma^m )}
\end{align*}
where 
\begin{align*}
    p_m = \frac{w_m\Tr(\sigma^m )}{\sum_{m = 1}^{M} w_m\Tr(\sigma^m )}.
\end{align*}
i.e. $F_{\vec{p}}$ is a weighted sum of $\frac{\Tr(\sigma_0  \sigma^m )}{\Tr(\sigma^m )}$. Therefore, we have:
\begin{align*}
    \max_{\vec{p}} F_{\vec{p}}&\leq \max_{m}  \frac{\Tr(\sigma_0  \sigma^m )}{\Tr(\sigma^m )}
\end{align*}
From Refs.~\cite{koczorExponentialErrorSuppression2020,hugginsVirtualDistillationQuantum2021} we see that:
\begin{align*}
    \max_{m}  \frac{\Tr(\sigma_0  \sigma^m )}{\Tr(\sigma^m )} =  \frac{\Tr(\sigma_0  \sigma^M )}{\Tr(\sigma^M )},
\end{align*}
which in terms implies the $\vec{w}$ that maximise $F_{\vec{p}}$ is the one with $w_M = 1$ and all other $w_m=0$.

Hence, the optimal symmetry-expanded density operator is:
\begin{align*}
    \frac{C_M \rho }{\Tr(C_M \rho)}
\end{align*}
which is just the density operator used for virtual distillation in Refs.~\cite{koczorExponentialErrorSuppression2020,hugginsVirtualDistillationQuantum2021}. Symmetry expansion in this case was shown to be able to reduce the infidelity exponentially with increase of $M$ while symmetry verification can only reduce the infidelity linearly with increase of $M$.

Besides the performance improvement, using symmetry expansion over symmetry verification in this case and more generally for non-abelian symmetry groups also has advantages in terms of circuit implementation. For symmetry expansion, we only require measuring one symmetry operator along with the observable in the same circuit run. If the expectation value of the symmetry operator is real, then no ancilla and complicated long-range gates between ancilla and data are required for symmetry expansion. In the example of virtual distillation, such circuits are constructed in Ref.~\cite{hugginsVirtualDistillationQuantum2021}.

\section{Fraction of Detectable Errors in Fermi-Hubbard Simulation}\label{sec:frac_of_detect_err}
Similar discussion can be found in Ref.~\cite{caiResourceEstimationQuantum2020}. Since the quantum states in between our two-qubit gates have well-defined $G_{tot}$, any Pauli errors that anti-commute with $G_{tot}$ will flip the $G_{tot}$ value and get detected. Similarly for $G_{\uparrow}$ and  $G_{\downarrow}$. Under Jordan-Wigner transformation, $G_{tot}$ is simply the tensor product of $Z$ acting on all qubits while $G_{\uparrow/\downarrow}$ is simply the tensor product of $Z$ acting on the qubits in the spin-up/down subspace.

\subsection{Depolarising Error}
Here we assume all of the two-qubit gates in the circuit are affected by two-qubit depolarising errors. Hence, there are 15 possible two-qubit Pauli errors after each two-qubit gates, only the error components generated by $\{Z_1, Z_2, X_1X_2\}$ (excluding identity) are undetectable by $G_{tot}$. Hence, $8$ out of $15$ Pauli error components are detectable by $G_{tot}$, which means
\begin{align*}
    f_{G_{tot}} = \frac{8}{15}.
\end{align*}

Now to calculate the fraction of errors detectable by $G_{\uparrow}$, we need to divide our two-qubit gates into three types:
\begin{itemize}
    \item Gates acting across both spin spaces:
    
    Suppose qubit $1$ is in spin up and qubit $2$ is in spin down, then the two-qubit error components that are undetectable by $G_{\uparrow}$ is generated by $\{Z_1, Z_2, X_2\}$, excluding identity. Hence, for the errors in these gates, the fraction of error detectable by $G_{\uparrow}$ is $\nicefrac{8}{15}$.
    
    \item Gates acting within the spin-up subspace:
    
    Similar to our discussion of $G_{tot}$, $\nicefrac{8}{15}$ of the errors are detectable by $G_{\uparrow}$
    
    \item Gates acting within the spin-down subspace:
    
    All errors are undetectable by $G_{\uparrow}$.
\end{itemize}
Since our circuit consists of alternating layers of gates acting across the spin subspaces and acting within the spin subspaces, we should expect roughly $\frac{1}{2}$ of the gates are acting across different spin spaces, roughly $\frac{1}{4}$ of the gates are acting within the spin-up subspace and roughly $\frac{1}{4}$ within the spin-down subspace. Hence, the total fraction of errors detectable by $G_{\uparrow}$ is thus:
\begin{align*}
    f_{G_{\uparrow}} \sim \frac{1}{2} \times \frac{8}{15} + \frac{1}{4} \times \frac{8}{15} + \frac{1}{4} \times 0   = \frac{2}{5}.
\end{align*}
Similarly we also have:
\begin{align*}
    f_{G_{\downarrow}} \sim \frac{2}{5}.
\end{align*}

\subsection{Bit-flip Error}\label{sec:bit_flip_fraction}
Here we assume all of the two-qubit gates in the circuit are followed by two single-qubit bit-flip error locations, i.e. after applying each two-qubit gate, we have $(1-p)^2$ probability that no error happens, $(1-p)p$ probability that one of the qubits is flipped and $p^2$ probability that both qubits are flipped. 
To consider the fraction of detectable errors, we only need to consider fraction of detectable errors at each error location (\emph{not} each gate), which in this case are all just single-qubit bit-flip error locations. Any single-qubit bit-flip in the circuit will anti-commute with $G_{tot}$ thus they are all detectable:
\begin{align*}
    f_{G_{tot}} = 1.
\end{align*}

Now to calculate the fraction of errors detectable by $G_{\uparrow}$, we simply need to note that half of our single-qubit bit-flip error locations are in the spin-up subspace, which produce errors that are detectable by $G_{\uparrow}$, while the other half if error locations are in the spin-down subspace, which produce errors that are undetectable by $G_{\uparrow}$. Hence, we have:
\begin{align*}
    f_{G_{\uparrow}} = \frac{1}{2}.
\end{align*}
and similarly:
\begin{align*}
    f_{G_{\downarrow}} = \frac{1}{2}.
\end{align*}

\section{Numerical Simulation for Bit-flip Noise}\label{sec:bit_flip_sim}
The same circuit as in \cref{sec:simulation} is used for our simulation, which consists of two-qubit components that are affecting by bit-flip noise as described in \cref{sec:bit_flip_fraction}. We have also shown in \cref{sec:bit_flip_fraction} that the fraction of errors detectable by $G_{tot}$ is estimated to be $f_{G_{tot}} \sim 1$ and that by $G_{\uparrow/\downarrow}$ is estimated to be $f_{G_{\uparrow/\downarrow}} \sim \frac{1}{2}$. Following \cref{eqn:opt_expand_f} and the discussions there, the small-bias expansion scheme is simply expanding with
\begin{align}\label{eqn:hubbard_opt_exp_bit}
    \mathbb{G}^* = \{G_{\uparrow/\downarrow}\}.
\end{align} 

\begin{figure*}[htbp]
    \centering
    \subfloat[8-Qubit]{\includegraphics[height = 0.32\textwidth]{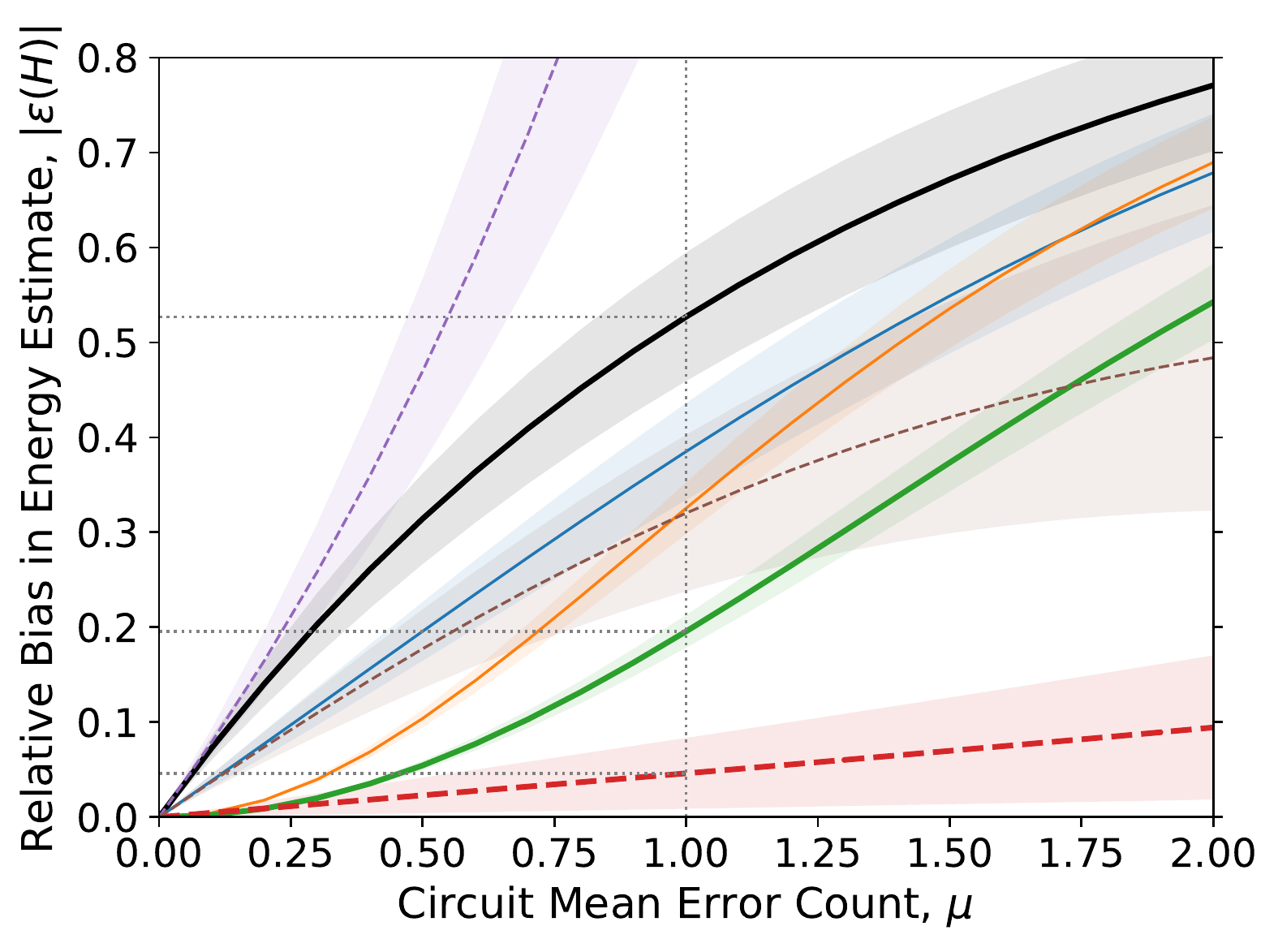}}\quad
    \subfloat[12-Qubit]{\includegraphics[height = 0.32\textwidth]{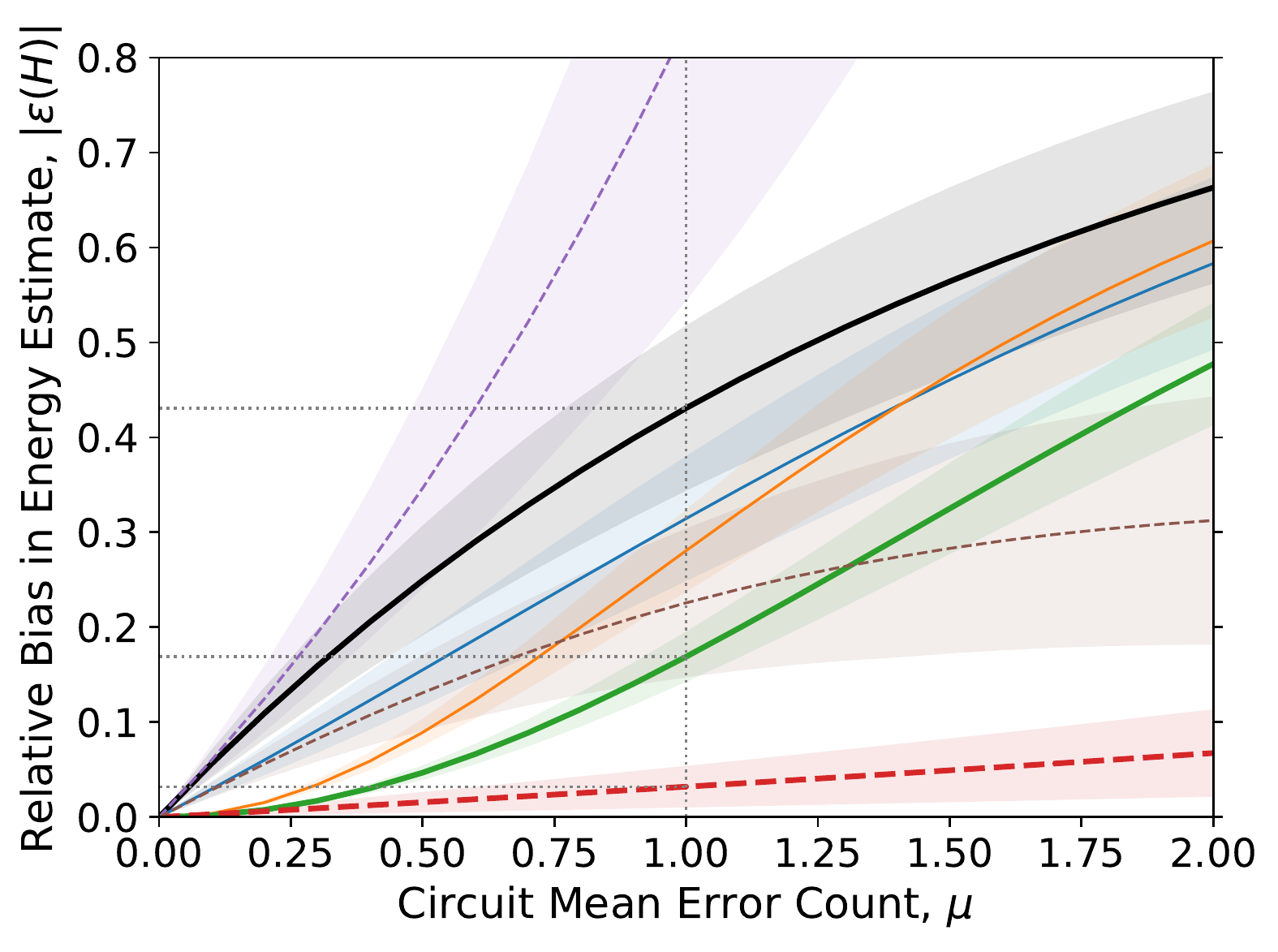}}\\
    \subfloat[8-Qubit]{\includegraphics[height = 0.32\textwidth]{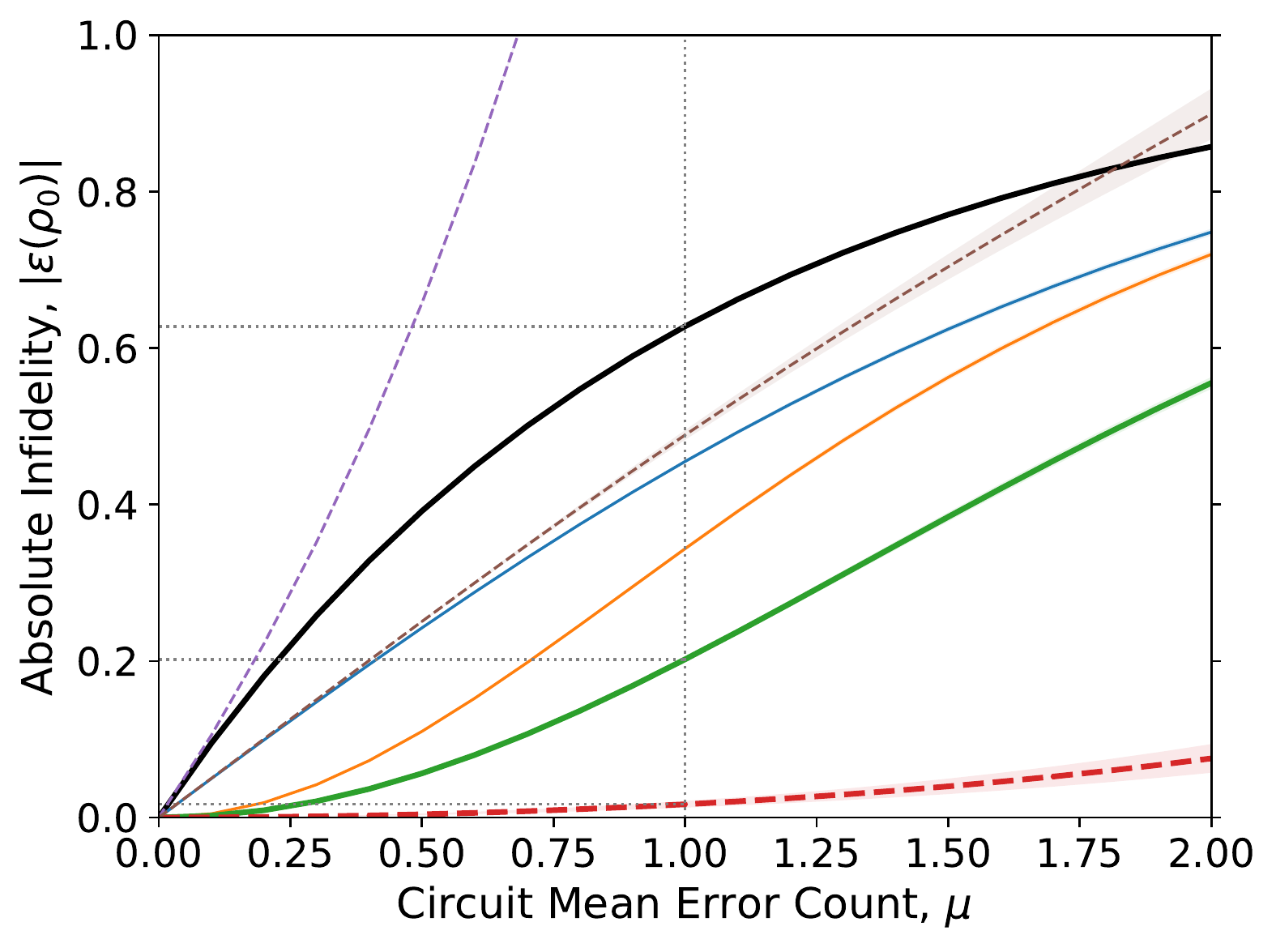}}\quad
    \subfloat[12-Qubit]{\includegraphics[height = 0.32\textwidth]{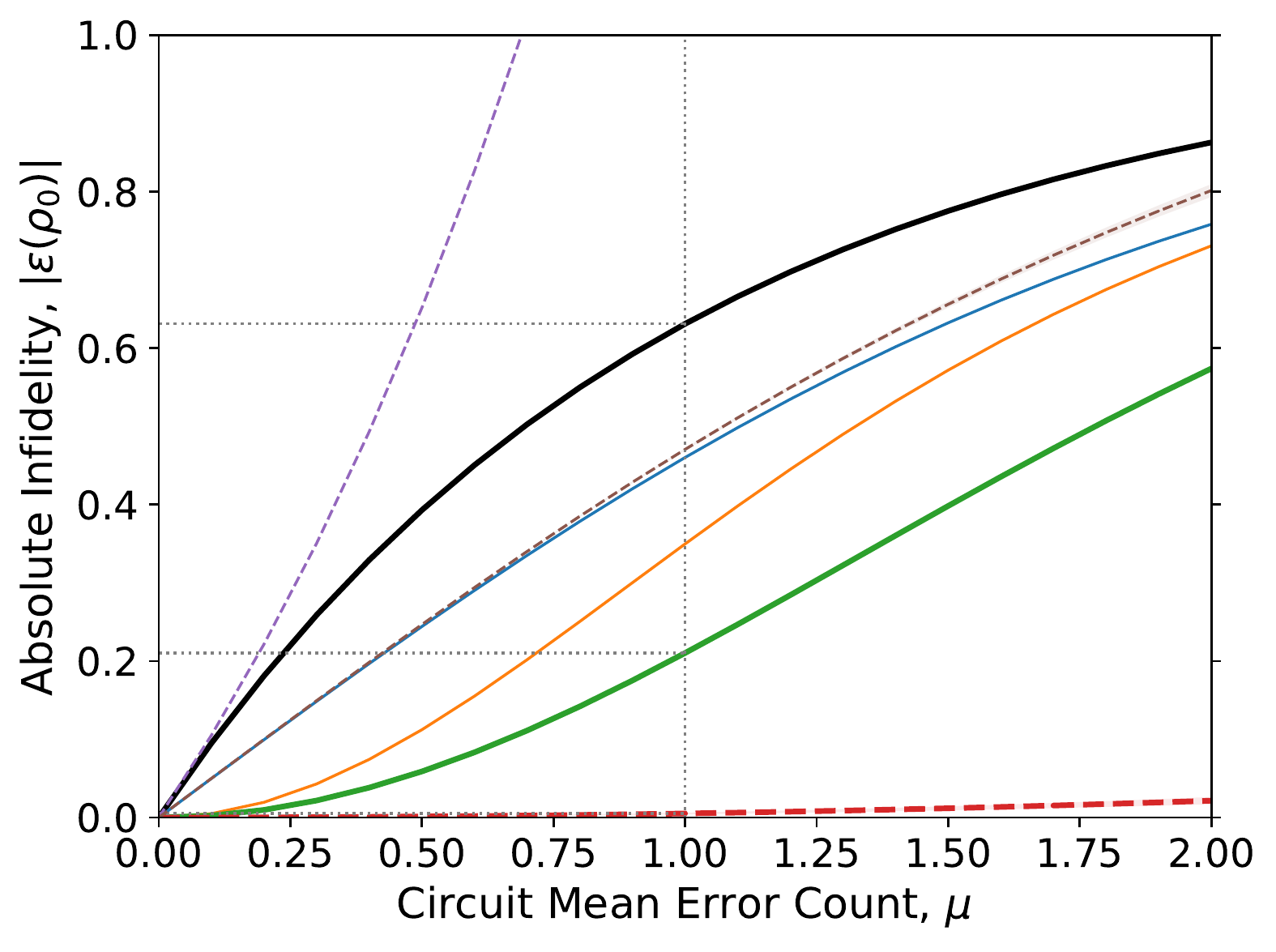}}\\
    \subfloat[]{\quad\quad\quad\quad \quad \includegraphics[height = 0.32\textwidth]{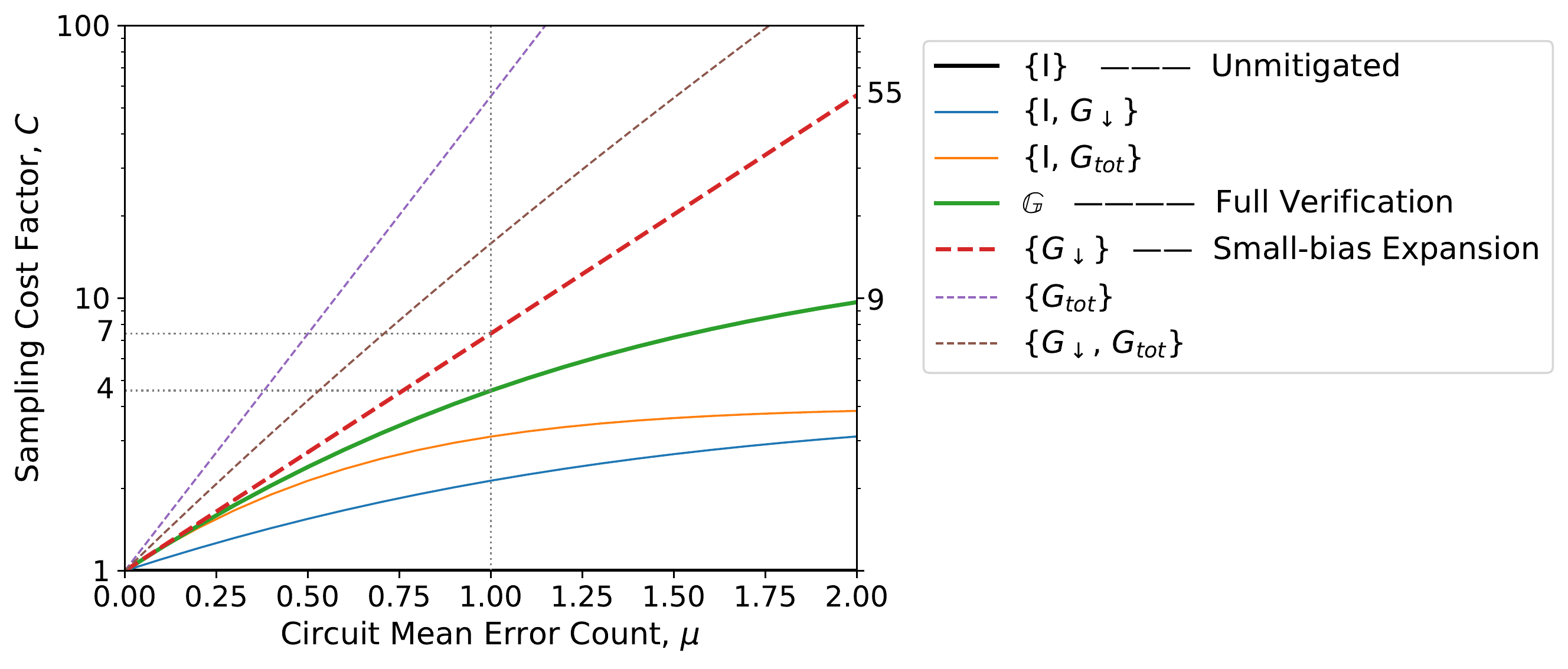}}
    \caption{Performance metrics for different symmetry expansion schemes against the increase of mean circuit error count under bit-flip noise: the relative bias in the energy estimate for (a) 8-qubit and (b) 12-qubit simulations;  the absolute infidelity for (c) 8-qubit and (d) 12-qubit simulations; (e) the sampling cost for all simulations. Data are averaged over circuit configurations with randomly generated parameters. The shaded areas indicate the spread of the lines over different parameters. The symmetry expansions we consider are all uniform expansion of a given set of symmetry operators. The legend indicates the set of symmetry operators used in the corresponding symmetry expansion. When the set of symmetry operators forms a group, the corresponding symmetry expansion will be symmetry verification which is labelled using solid lines. All the other symmetry expansion are labelled using dashed lines. Three of the most representative methods: without mitigation (black), symmetry verification using the full symmetry group $\mathbb{G} = \{I,\ G_{\uparrow},\ G_{\downarrow},\ G_{tot}\} $ (green) and the small-bias symmetry expansion using  $\mathbb{G}^* = \{G_{\downarrow}\}$ (red)  are labelled with thicken lines.}
    \label{fig:sim_result_bit}
\end{figure*}

The absolute relative bias in the energy estimate over different mean circuit error counts $\mu$ for the $8$-qubit and $12$-qubit simulations are shown in \cref{fig:sim_result_bit} (a) and (b). The corresponding absolute infidelity $\abs{\epsilon(\rho_0)}$ are shown in \cref{fig:sim_result_bit} (c) and (d). The corresponding sampling costs are shown in \cref{fig:sim_result_bit} (e). In \cref{fig:sim_result_bit}, we again see that out of all symmetry verification schemes (solid lines), the symmetry verification using the full symmetry group $\mathbb{G}$ can achieve the lowest bias. When we look at all symmetry expansion schemes including symmetry verifications, we see that the small-bias scheme we found again can achieve the lowest bias. The bias and sampling cost comparison among the small-bias scheme, symmetry verification and without mitigation is shown in \cref{tab:est_err_bit}. At $\mu = 1$, the small-bias scheme can achieve a much smaller bias compared to symmetry verification and without mitigation at a moderate sampling cost. At $\mu = 2$, the small-bias scheme can maintain the small bias, but at a larger sampling cost. This is similar to what happens for the depolarising noise in the main text.

\begin{table}[t]
    \centering
    \begingroup
    \renewcommand{\arraystretch}{1.5}
    \setlength{\tabcolsep}{-1pt}
    \subfloat[$\mu=1$ ]{
        \begin{tabular}{lccccccc}\toprule
            & \multicolumn{2}{c}{8-Qubit} & \phantom{abc}& \multicolumn{2}{c}{12-Qubit}& \phantom{abc} & \multirow{2}{*}{Cost}\\
            \cline{2-3} \cline{5-6}
            & $\ \abs{\epsilon(H)}\ $ & $\ \abs{\epsilon(\rho_0)}\ $& &  $\ \abs{\epsilon(H)}\ $ & $\ \abs{\epsilon(\rho_0)}\ $ && \\ \hline
            Unmitigated & 0.527 & 0.628 &     & 0.431 & 0.631 &     &  1.0 \\
            Verified    & 0.196 & 0.202 &     & 0.169 & 0.210 &     &  4.6 \\
            Expanded    & 0.046 & 0.017 &     & 0.032 & 0.005 &     &  7.4 \\
            \botrule
    \end{tabular}}\\
    \subfloat[$\mu=2$ ]{
        \begin{tabular}{lccccccc}\toprule
            & \multicolumn{2}{c}{8-Qubit} & \phantom{abc}& \multicolumn{2}{c}{12-Qubit}& \phantom{abc} & \multirow{2}{*}{Cost}\\
            \cline{2-3} \cline{5-6}
            & $\ \abs{\epsilon(H)}\ $ & $\ \abs{\epsilon(\rho_0)}\ $& &  $\ \abs{\epsilon(H)}\ $ & $\ \abs{\epsilon(\rho_0)}\ $ && \\ \hline
            Unmitigated & 0.771 & 0.857 &     & 0.663 & 0.863 &     &  1.0 \\
            Verified    & 0.543 & 0.556 &     & 0.477 & 0.574 &     &  9.7 \\
            Expanded    & 0.094 & 0.076 &     & 0.068 & 0.022 &     & 55.6 \\
            \botrule
    \end{tabular}}
    \endgroup
    \caption{The relative bias in energy estimate $\abs{\epsilon(H)}$,  absolute infidelity $\abs{\epsilon(\rho_0)}$, and the sampling costs for three of the most representative symmetry expansions in our simulation: the unmitigated noisy state (Unmitigated), symmetry verification using the full symmetry group $\mathbb{G}$ (Verified) and the small-bias symmetry expansion using  $\{G_{\downarrow}\}$ (Expanded) at the mean circuit error rate (a) $\mu = 1$ and (b) $\mu = 2$. Note that the cost for symmetry verification here are for post-processing symmetry verification. }
    \label{tab:est_err_bit}
\end{table}
    
\newpage

\end{document}